\documentclass[journal]{IEEEtran}
\usepackage{amsmath,amssymb,enumerate,graphicx}
\usepackage[dvipsnames]{xcolor}
\usepackage[mathscr]{eucal}

\usepackage{float}
\usepackage{graphicx} \usepackage{siunitx}
\usepackage{booktabs}
\usepackage{hyperref}
\usepackage{subcaption}

\newcommand{\br}{\mathbb{R}}
\newcommand{\trs}{^{\top}}
\newcommand{\qed}{\hfill $\Box$}
\newcommand{\stack}{\mbox{stack}}
\newcommand{\ve}{\varepsilon}
\newcommand{\p}{\prime}
\newcommand{\pp}{{\prime\prime}}
\newcommand{\ov}[1]{\overline{#1}}
\newtheorem{assumption}{Assumption}[section]
\newtheorem{proposition}{Proposition}[section]
\newtheorem{definition}{Definition}[section]
\newtheorem{lemma}{Lemma}[section]
\newtheorem{corollary}{Corollary}[section]

\newtheorem{remark}{Remark}[section]

\def\diag{\mbox{diag}}

\def\rea{\mathbb{R}}

\def\L2e{{\cal L}_{2e}}

\def\qed{\hfill$\Box \Box \Box$}

\def\begequarr{\begin{eqnarray}}
\def\endequarr{\end{eqnarray}}
\def\begequarrs{\begin{eqnarray*}}
\def\endequarrs{\end{eqnarray*}}
\def\begarr{\begin{array}}
\def\endarr{\end{array}}
\def\begequ{\begin{equation}}
\def\endequ{\end{equation}}
\def\lab{\label}
\def\begdes{\begin{description}}
\def\enddes{\end{description}}
\def\begenu{\begin{enumerate}}
\def\begite{\begin{itemize}}
\def\endite{\end{itemize}}
\def\endenu{\end{enumerate}}

\def\lef[{\left[\begin{array}}
\def\rig]{\end{array}\right]}
\def\begcen{\begin{center}}
\def\endcen{\end{center}}
\def\begrem{\begin{remark}\rm}
\def\endrem{\end{remark}}

\def\cale{{\cal E}}

\def\cale{{\cal E}}

\def\calk{{\cal K}}

\def\begmat#1{\begin{bmatrix}#1\end{bmatrix}}

\usepackage{color}

\usepackage[prependcaption,colorinlistoftodos]{todonotes}

\begin{document}

\title{On the Existence and Long-Term Stability of Voltage Equilibria in Power Systems with Constant Power Loads}
\author{Alexey S. Matveev, Juan E. Machado, Romeo Ortega,~\IEEEmembership{Fellow,~IEEE}, Johannes Schiffer and Anton Pyrkin, ~\IEEEmembership{Member,~IEEE}%
\thanks{Alexey S. Matveev is with  the Department of Mathematics and Mechanics, Saint Petersburg State University, St. Petersburg 198504, Russia. E-mail: almat1712@yahoo.com.}%
\thanks{Juan E. Machado and Romeo Ortega are with Laboratoire des Signaux et Syst\`emes (L2S), CentraleSup\'elec, 91192, Gif-Sur-Yvette, France. E-mail: juan.machado@l2s.centralesupelec.fr (Juan E. Machado), romeo.ortega@lss.supelec.fr (Romeo Ortega).}%
\thanks{Johannes Schiffer is with Control Systems and Network Control Technology, Brandenburg University of Technology (BTU), 03046, Cottbus, Germany. Email: schiffer@b-tu.de.}%
\thanks{Anton Pyrkin is with the Department of Control Systems
and Informatics, ITMO University, Saint Petersburg 197101, Russia. E-mail: a.pyrkin@gmail.com.}}%

\markboth{Submitted to IEEE Transactions on Automatic Control. This version: SEPTEMBER 21, 2018}%
{}


\maketitle
%
%
\begin{abstract}
Voltage instability is a major threat in power system operation. The growing presence of constant power loads significantly aggravates this issue, hence motivating
the development of new analysis methods for both existence and stability of voltage equilibria. 
Formally, this problem can be cast as the analysis of solutions of a set of nonlinear algebraic equations of the form $f(x)=0$, where $f:\rea^n \mapsto \rea^n$, and the associated differential equation $\dot x=f(x)$. By invoking advanced concepts of dynamical systems theory and effectively exploiting its monotonicity, we exhibit all possible scenarios for existence, uniqueness and stability, of its equilibria. We prove that, if there are equilibria, there is a distinguished one that is locally stable and attractive, and we give some physically-interpretable conditions such that it is unique. Moreover, a simple on-line procedure to decide whether equilibria exist of not, and to compute the distinguished one is proposed.  
In addition, we show how the proposed framework can be applied to 
long-term voltage stability analysis in AC power systems, multi-terminal high-voltage DC systems and DC microgrids.
 
\end{abstract}


%
\section{Introduction}
\lab{sec1}
%

A {\em sine qua non} condition for the correct operation of power systems is the existence of a steady-state behavior that, moreover, should be robust in the presence of perturbations \cite{kundur94}. Viewed as dynamical systems, described with differential equations, this requirement translates into the existence  of equilibria, which should also be stable and attractive. The accurate description of modern power systems necessarily incorporates ``strong" nonlinear effects, complicating the task of analysis of its equilibria. 

Variables of particular importance in both AC and DC power systems are the voltage magnitudes at the different nodes of the system. In fact, during the past decades an increasing number of incidents can be attributed to fast and slow voltage variations \cite{kundur04,vanCutsem07}. Hence, voltage stability analysis has significantly gained in relevance in AC power systems \cite{kundur94,kundur04,vanCutsem07,simpson16}. In DC power systems the voltage magnitudes can be considered even more relevant, since---in the absence of a system frequeny---variations in the system loading always have a direct impact on the DC voltages \cite{jovcic11}. 


In this paper we derive a methodological approach, which permits to determine existence and stability properties of voltage equilibria in a broad range of power system applications. More precisely, we show that our proposed approach is applicable to analyze the steady-state voltage behavior of traditional AC power systems \cite{kundur94,vanCutsem07} as well as of two emerging power system concepts, namely multi-terminal high-voltage (MT-HV) DC networks \cite{vanHertem10,jovcic11} and DC microgrids \cite{elsayed15,dragivcevic16}. 

In addition, if stationary voltage solutions exist our method also allows to identify the solution with the highest voltage magnitudes as well as to assert its long-term stability properties. Following standard practice \cite{lof93,venikov75,venikov77,kundur94,hill93}, the latter notion is defined in terms of the eigenvalues of the Jacobian of the algebraic power system equations evaluated at a stationary solution.

In all the examples mentioned above, the key problem is the study of a nonlinear algebraic equation $f(x)=0 \in \rea^n$ in $x \in \rea^n$, where only solutions $x$ with positive components are of interest. The approach adopted in the paper to tackle these problems is to associate to $f(x)$ the ordinary differential equation (ODE) $\dot x =f(x)$, and to apply to it tools of dynamical systems \cite{Hart82} to study existence and stability of its equilibria, which are nothing but the solutions of the primal algebraic equation.

The main contributions of our work are the proofs of the following properties of the ODE.

\begenu[{\bf C1.}]
\item If there are no equilibria (stable or unstable) then, in all solutions of the ODE, one or more components converge to zero in finite time.
\item If equilibria exist, there is a distinguished equilibrium, say $\bar x_{\tt max}$, among them that dominates component-wise all the other ones. This equilibrium $\bar x_{\tt max}$ is locally  stable and attracts all trajectories that start in a certain well-defined domain. 
\item By solving a system of $n$ convex algebraic inequalities in $n$ positive unknowns we explicitly identify a set of initial states with the following characteristics: (i) all trajectories starting there monotonically decay in all components; (ii) they either  have at least one component that converges to zero in finite time or none of them does. Moreover, in the latter case, the trajectory is forward complete and converges to $\bar x_{\tt max}$.
\endenu

Clearly, the contribution {\bf C3} suggests a simple on-line computational procedure to answer the questions raised in the paper: find some solution of the convex inequalities mentioned in {\bf C3}, run a simulation of $\dot x=f(x)$ starting from this set, and check whether there is a component of the trajectory that converges to zero in finite time and, if not, find the limit state $\bar x_{\tt max}$ of the trajectory, which is an asymptotically stable equilibrium. An additional contribution is to give physically-interpretable conditions on the problem data that ensure $\bar x_{\tt max}$ is the only stable equilibrium.

The remainder of the paper is organized as follows. Section \ref{sec2} describes the ODE $\dot x=f(x)$ of interest and gives the main theoretical results pertaining to it. In  Section \ref{sec3} we illustrate these results with three canonical power systems examples. Section \ref{sec4} presents some numerical simulation results. The paper is wrapped-up with concluding remarks in  Section \ref{sec5}. To enhance readability, all proofs of the technical results are given in Appendices at the end of the paper.\\      

\noindent {\bf Notation} $(\cdot)\trs$ denotes transposition, $\br$ is the real line, $\br^n$ is the Euclidean space of vector columns $x = (x_1, \ldots, x_n)\trs$, its positive orthant is denoted as $\calk_+^n := \{x \in \br^n: x>0\}$, $\stack (p_i) \in \br^{r_1+\cdots+r_N}$, denote stacking $p_i \in \br^{r_i}, i \in \{1,\dots,N\}$ on top of one another, $\diag (A_1, \ldots, A_k)$, is the block-diagonal matrix composed of the listed square blocks $A_i$. Inequalities between vectors $x,y \in \br^n$ are meant component-wise. All mappings are assumed smooth. Given a mapping $f:\rea^n\to \rea^n$ we denote its Jacobian by $\nabla f(x):={\partial f(x) \over \partial x}$. The operator $\langle \cdot \rangle $ denotes the clipping function $\langle a \rangle = \max\{a,0\}$.
%
\section{Analysis of the ODE of Interest}
\lab{sec2}
%
As indicated in the introduction, in this paper we are interested in the steady-state voltage solutions of AC power systems (under the common decoupling assumption \cite{kundur94}), MT-HVDC networks as well as DC microgrids.
In Section \ref{sec3} it is shown that this study boils down to the analysis of solutions of the following algebraic equation
\begin{equation}
\label{stestaequ}
A \bar x + \stack\left( \frac{b_i}{\bar x_i}\right) - w =0
\end{equation}
where $\bar x \in \calk_+^n$. Here $A \in \rea^{n \times n}$,  $b_i \in \br$, and $w \in \br^n$ are given and satisfy the following.

\begin{assumption}\em
\label{ass.1}
The matrix $A$ is symmetric and positive definite, all its off-diagonal elements are non-positive and $b_i \neq 0$ for all $i$.
\end{assumption}

To study the solutions of \eqref{stestaequ} we consider the following ODE
\begin{equation}
\label{eq.1}
\dot{x} = f(x):= - A x - \stack\left( \frac{b_i}{x_i}\right) + w,
\end{equation}
and we are interested in studying the existence, and stability, of the equilibria of \eqref{eq.1}. In  particular, we will provide answers to the following questions.
\begenu[{\bf Q1}]
\item When do equilibria exist? Is it possible to offer a simple test to establish their existence?
\item If there are equilibria, is there a distinguished element among them?
\item Is this equilibrium stable and/or attractive?
\item  If it is attractive, can we estimate its domain of attraction?
\item Is it possible to propose a simple procedure to compute this special equilibrium using the system data $(A,b,w)$?
\item Are there other stable equilibria?
\endenu

Instrumental to provide answers to the questions {\bf Q1---Q6} is the fact that the system \eqref{eq.1} is {\em monotone}. That is, for any two solutions $x_a(\cdot),x_b(\cdot)$ of \eqref{eq.1}, defined on a common interval $[0,T]$, the inequality $x_a(0) \leq x_b(0)$ implies that $x_a(t) \leq x_b(t)\; \forall t \in [0,T]$. This can be easily verified by noticing that equation \eqref{eq.1} satisfies the necessary and sufficient condition for monotonicity \cite[Proposition 1.1 and Remark 1.1, Ch. III]{SMI}
$$
{\partial f_i(x) \over \partial x_j} \geq 0,\;\forall i \neq j,\;\forall x \in \calk_+^n.
$$

In the sequel, we denote by $x(t,x_0)$ the solution of \eqref{eq.1} with initial conditions $x(0)=x_0>0$, and use the following.

\begin{definition}\em
\label{def.prpr}
An equilibrium $\bar x >0$ of \eqref{eq.1} is said to be {\em globally attractive from the right} if for any $x_0 \geq \bar x$, the solution $x(t,x_0)$ is defined on $[0,\infty)$ and converges to $\bar x$ as $t \to \infty$. The equilibrium is said to be {\em hyperbolic} if the Jacobian matrix $\nabla f(\bar x)$ has no eigenvalue with zero real part \cite{Hart82}.
\end{definition}

%
%
\subsection{The simplest example}
\begin{figure}
\centering
\begin{subfigure}[b]{0.25\textwidth}
\includegraphics{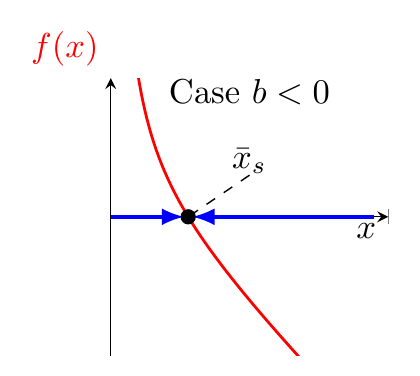}
\caption{}
\end{subfigure}%
~
\begin{subfigure}[b]{0.25\textwidth}
\includegraphics{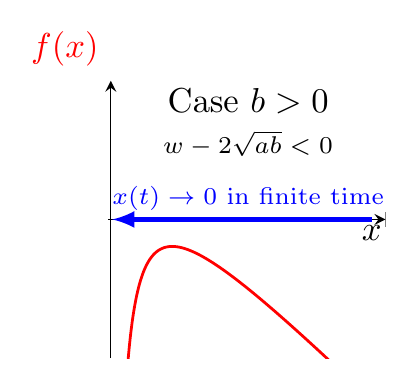}
\caption{}
\end{subfigure}\\
\begin{subfigure}[b]{0.25\textwidth}
\includegraphics{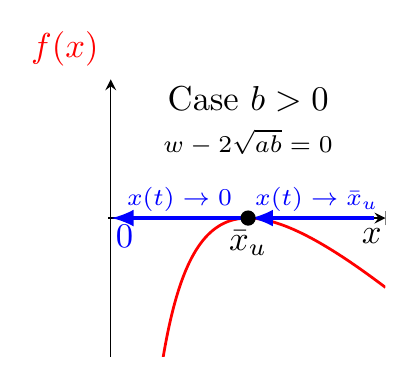}
\caption{}
\end{subfigure}%
~
\begin{subfigure}[b]{0.25\textwidth}
\includegraphics{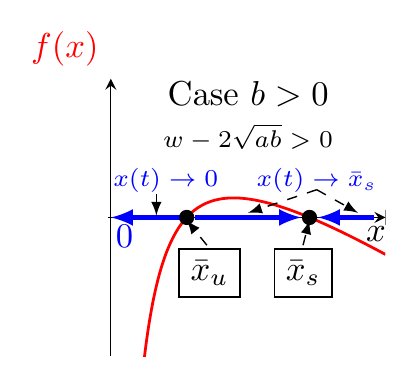}
\caption{}
\end{subfigure}%
~
\caption{ Feasible behaviors of the one-dimensional system \eqref{onedimsys}: (a) A unique globally attractive equilibrium $\bar x_{s}$; (b) No equilibria, all solutions converge to zero in a finite time $t_f$; (c) Unique unstable equilibrium $\bar x_u$, which is globally attractive from the right, whereas any solution starting on the left diverges from $\bar x_u$ and converges to $0$ in a finite time; (d) Two equilibria, the smallest of which $\bar x_u$ is unstable, whereas the larger one $\bar x_s$ is locally stable and globally attractive from the right.}
\label{fig.1}
\end{figure}

\lab{subsec21}
To gain an understanding of some key traits of possible results, it is instructive to start with the simplest case $n=1$. Then, $x \in \br$ and \eqref{eq.1} is the scalar equation
\begequ
\lab{onedimsys}
\dot{x} = - a x - \frac{b}{x} + w,
\endequ
where $a>0, b \neq 0$. Feasible behaviors of the system are exhaustively described in Figure~\ref{fig.1}.

The following can easily be inferred from this figure:
\begin{enumerate}[{\bf p.1)}]
\item The system has no equilibria, it has finitely many equilibria, or a single equilibrium.
\item If the system has equilibria, the rightmost of them it is globally attractive from the right.
\item Non-hyperbolic equilibria may be globally attractive from the right but are not locally stable; apart from such equilibria, there may be no other ones.
\item Hyperbolic and globally attractive from the right equilibria are locally stable.
\item If $b > 0$, globally stable equilibria do not exist.
\end{enumerate}

We will show below  that several of the traits mentioned above are inherited by the $n$-th order ODE \eqref{eq.1}. 
\subsection{A generic assumption}
\lab{subsec22}
%
Situation p.3) above is, clearly, undesirable. Since this can happen in the general case---{\em e.g.}, considering a diagonal matrix $A$---it is reasonable to exclude its possible appearance.  

\begin{assumption}\em
\label{ass.2}
There are {\em no non-hyperbolic equilibria} of the system \eqref{eq.1}. This is, clearly, equivalent to assuming that the following set identity holds
{\footnotesize
\begin{equation}
\label{infeas}
\left\{x \in \calk_+^n\;|\;\det \left[ A - \diag\left( \frac{b_i}{x_i^2}\right) \right] =0 , \; w = Ax + \stack\left( \frac{b_i}{x_i}\right)\right\}=\emptyset.
\end{equation}
}
\label{ass.zero}
\vspace{-5mm}
\qed
\end{assumption}

The lemma below proves that Assumption \ref{ass.2} is almost surely true, hence it is done without loss of generality. The proof of the lemma is given in Appendix \ref{appa}.

\begin{lemma}\em
\label{lem.singular}
For any given $A$ and $b_i \neq 0$, the set of all $w \in \br^n$ for which Assumption~{\rm \ref{ass.zero}} does not hold has zero Lebesgue measure and is nowhere dense.
\qed
\end{lemma}

\subsection{Main results on system \eqref{eq.1}}
\label{sec.mres}
%
The first proposition contains a qualitative analysis of the system. 

\begin{proposition}\em
\label{th.main0}
Consider the system \eqref{eq.1} verifying Assumptions~{\rm \ref{ass.1}} and {\rm \ref{ass.2}}. One and only one of the following two mutually exclusive statements holds.
\begin{enumerate}[{\bf s.1)}]
\item There are no equilibria $\bar x$, either stable or unstable, and any solution $x(\cdot)$ is defined only on a finite time interval $[0,t_f) \subset [0,\infty)$, since for any of them, there exists at least one coordinate $x_i$ such that $x_i(t) \to 0, \dot{x}_i(t) \to - \infty$ as $t \to t_f$. Such a coordinate is necessarily associated with $b_i >0$.\footnote{So the case {\bf s.1)} does not occur if $b_j < 0$,  $\forall~j$.}
\item There exist one or finitely many equilibria $\bar x^k$. One of them $\bar x_{\tt max}>0$ verifies $\bar x_{\tt max} \geq \bar x^k,\; \forall k$, and this equilibrium is locally stable and attractive from the right.
\end{enumerate}

If all $b_i$'s are of the same sign, then in the case {\bf s.2)}, there are no other locally stable equilibria apart from $\bar x_{\tt max}$.
\qed
\end{proposition}

The proof of this proposition is given in Appendix~\ref{appc}.\\

The next proposition provides a constructive test to identify which of the cases s.1) or s.2) holds, as well as a method to find $\bar x_{\tt max}$ in the case s.2). To articulate the result, we introduce the following.

\begin{definition}\em
A solution $x(\cdot)$ of the differential equation \eqref{eq.1} is said to be {\em characteristic} if its initial condition lives in the set
\begequ
\lab{in.ch}
\cale:=\left\{ x \in \calk_+^n\; |\;A x > \stack \left( \langle w_i \rangle +\frac{\langle -b_i \rangle}{x_i}\right)\right\}.
\endequ

If all coefficients $b_i>0 \; \forall i$, the set \eqref{in.ch} reduces to the (convex open polyhedral) cone $\left\{ x \in \calk_+^n\; |\;A x >\stack \left( \langle w_i \rangle \right)  \right\}$.
\qed
\end{definition}

\begin{proposition}\em
\label{th.main}
Consider the system \eqref{eq.1} verifying Assumptions~{\rm \ref{ass.1}} and {\rm \ref{ass.2}}.
\begin{enumerate}[{\bf I)}]
\item The set $\cale$ is non-empty, consequently there are characteristic solutions.
\item All characteristic solutions $x(\cdot)$ strictly decay, in the sense that $\dot{x}(t) <0$, for all $t$ in the domain of definition of $x(\cdot)$.
\item One and only one of the following two mutually exclusive statements holds for all characteristic solutions $x(\cdot)$:
\begin{enumerate}[{\bf (i)}]
\item For a finite time $t_f \in (0,\infty)$, some coordinate $x_i(\cdot)$ approaches zero:
\begin{equation}
\label{conv.zero}
x_i(t) \to 0 \qquad \text{\rm as} \quad t \to t_f,
\end{equation}
and the solution $x(\cdot)$ is defined only on the finite time interval $[0,t_f)$.
\item There is no coordinate approaching zero, the solution is defined on $[0,\infty)$, and the following limit exists and verifies
\begin{equation}
\label{limt}
\lim_{t \to \infty} x(t) >0.
\end{equation}
This limit is the same for all characteristic solutions.
\end{enumerate}
\end{enumerate}

\begin{itemize}
\item[{\bf IV)}] If the case {\bf (i)} holds for a characteristic solution, the situation {\bf s.1)} from Proposition~{\rm \ref{th.main0}} occurs.
\item[{\bf V)}] If the case {\bf (ii)} holds for a characteristic solution, the situation {\bf s.2)} from Proposition~{\rm \ref{th.main0}} occurs, and the dominant equilibrium $\bar x_{\tt max}$ is equal to the limit \eqref{limt}.
\end{itemize}
\qed
\end{proposition}

The proof of this proposition is given in Appendix \ref{appc}. 
%
\subsection{A procedure to verify Propositions \ref{th.main0} and \ref{th.main}}
\label{subsec24}
%
Proposition~\ref{th.main} suggests a computational procedure to verify whether the system has equilibria and, if they do exist, to find the dominant one $\bar x_{\tt max}$ among them, which is necessarily stable (and is the only stable equilibrium if all $b_i$'s are of the same sign). Specifically, it suffices to find an element of the set $\cale$  defined in \eqref{in.ch}, to launch the solution of the differential equation \eqref{eq.1} from this vector, and to check whether---as the solution decays---there is a coordinate approaching zero or, conversely, all of them remain separated from zero. In the last case, the solution will have a limit, which is precisely the stable equilibrium of the system.

The statement I of Proposition~\ref{th.main} ensures that the first step of this algorithm, {\em i.e.} generating an element of the set $\cale$ defined in \eqref{in.ch}), is feasible. Technically, this step consists in solving the following system of feasible {convex} inequalities:
$$
\langle w_i \rangle +\frac{\langle - b_i \rangle}{x_i} - \sum_{j=1}^n a_{ij} x_j < 0, \quad \forall i.
$$
This problem falls within the area of convex programming and so there is an armamentarium of effective tools to solve it. Nevertheless, this problem can be further simplified via transition from nonlinear convex inequalities to linear ones,  modulo closed-form solution of finitely many scalar quadratic equations. The basis for this is given by the following lemma, whose proof is given in Appendix \ref{appd}.

\begin{lemma}\em
\label{linear.lemma}
Pick any vector $z$ in the set $\left\{ x \in \calk_+^n\; |\;A x > 0\right\}.$\footnote{In Appendix \ref{appd} it is shown that, under Assumption~{\rm \ref{ass.1}}, this system of linear inequalities is feasible.} Define the scaled vector $x:=\mu z$, where
\begequ
\lab{mu}
\mu > \frac{\langle w_i \rangle + \sqrt{\langle w_i \rangle^2+ 4 (Az)_i \frac{\langle - b_i \rangle}{z_i}}}{2 (Az)_i},\;\forall i.
\endequ
Then, $x \in \cale$.
\qed
\end{lemma}
%
\subsection{Some additional properties of  system \eqref{eq.1}}
\label{subsec25}
%
\noindent {\bf P1} In {\bf III.i)}, there may be several coordinates $x_i$ with the described property, all coordinates do not necessarily  possess it, and different solutions $x(\cdot)$ may have distinct sets of coordinates with this trait.\\

\noindent {\bf P2} The claim {\bf s.1} in Proposition~\ref{th.main0} and {\bf IV} in Proposition~\ref{th.main} yield that \eqref{conv.zero} is necessarily associated with $b_i>0$ and $\dot{x}_i(t) \to - \infty$ as $t \to t_f$.\\ 

\noindent {\bf P3} Regarding the claim {\bf s.2} in Proposition~\ref{th.main0} the basin of attraction of the equilibrium $\bar x_{\tt max}$ is open and has the property that it contains all states $x \geq  \bar x_{\tt max}$.\\ 

\noindent {\bf P4} The linear programming problem of finding elements in the set $\left\{ x \in \calk_+^n\; |\;A x > 0\right\}$ has been widely studied in the literature \cite{Dantzig66,Schr98,BerTs97}. There is a whole variety of computationally efficient methods to solve this problem, including the Fourier-Motzkin elimination, the simplex method, interior-point/barrier-like approaches, and many others; for a recent survey, we refer the reader to \cite{DrMeWe18}.\\

\noindent {\bf P5} For any $i$ with $b_i>0$, the inequality \eqref{mu} clearly simplifies into   
$$
\mu > \frac{\langle w_i \rangle}{(Az)_i}.
$$ 
%
\section{Long-Term Voltage Stability Analysis of Some Canonical Power Systems}
\lab{sec3}
%
In this section we apply the results of Section \ref{sec2} to three different types of power systems. These comprise standard conventional AC power systems as well as MT-HVDC networks and DC microgrids---two promising emerging power system concepts.
These dynamical systems admit equilibrium points satisfying algebraic constraints that, under standard assumptions, can be written in the form \eqref{stestaequ} and verifying Assumption \ref{ass.1}. 
This permits the use of Propositions \ref{th.main0} and \ref{th.main} to study the existence and stability of equilibrium points. Moreover, we can also try the numerical procedure proposed in Subsection \ref{subsec24} to verify the claims of the propositions.

In all these examples, $x$ represents the vector of voltage magnitudes of the system. Following standard definitions and classifications of voltage stability in AC power systems \cite{lof93,venikov75,venikov77,hill93,kundur94,kundur04}, we introduce the following notion of {\it long-term voltage stability} for the system \eqref{stestaequ}, which relates the objectives stated above to standard power system practice. 

\begin{definition}\em
	A positive root $\bar x$ of the system \eqref{stestaequ} is long-term voltage stable if the Jacobian $\nabla f(x)\big|_{x=\bar x}$, with $f$ given in \eqref{eq.1}, is Hurwitz, {\em i.e.,} all its eigenvalues have a negative real part.
	\label{def:lvs}
\end{definition}

Definition~\ref{def:lvs} originates from a sensitivity analysis of the voltage magnitudes with respect to changes in the reactive power flows in AC networks, see \cite{venikov75,hill93,kundur94} and the more recent work \cite{simpson16}.

Lemma~\ref{lem.locst} in the Appendix implies that the Jacobian of the dynamics \eqref{eq.1} evaluated at any stable equilibrium point is Hurwitz. Hence, if case V) of Proposition~\ref{th.main} applies then the dominant equilibrium is long-term voltage stable in the sense of Definition~\ref{def:lvs}. 
Consequently, Proposition~\ref{th.main} provides a constructive procedure to evaluate the existence of a unique dominant and long-term stable voltage solution in power systems with constant power loads.

\subsection{Long-term voltage stability in AC power systems}
\lab{subsec33}
%
Consider a high-voltage AC power network with $n\geq 1$ nodes.
Denote by $V_i>0$ and $Q_i$ the voltage and the reactive power load demand at the node $i$, respectively. 
Under the standard decoupling assumption \cite{kundur94}, for each $i=1,...,n$, the decoupled reactive power flow, is given by \cite{kundur94,simpson16,schiffer16}
\begin{equation*}
Q_{\text{ZIP},i}=V_i \sum_{j=1}^n |B_{ij}|(V_i-V_k),
\end{equation*} 
where $B_{ij}<0$ if nodes $i$ and $j$ are connected via a power line and $B_{ij}=0$ otherwise. The reactive power demand $Q_{\text{ZIP},i}$ at the $i$-th node is  described by a, so-called,  ZIP model, {\it i.e.,} 
\begin{equation*}
Q_{\text{ZIP},i}:=\left(\mathcal{Y}_iV_i^2 + k_i V_i + Q_i \right).
\end{equation*}
The term ZIP load refers to a parallel connection of a constant impedance $\mathcal{Y}_i\in\mathbb{R}$, a constant current $k_i\in\mathbb{R}$, and a constant power $Q_i\in\mathbb{R}$ load. Then, we obtain the (algebraic) reactive power balance equation
\begin{equation}
\left(\mathcal{Y}_iV_i^2 + k_i V_i + Q_i \right)=V_i \sum_{j=1}^n |B_{ij}|(V_i-V_j),~i=1,...,n,
\label{reacpowbal}
\end{equation}
which by defining $x:=\stack\left(V_i\right)\in \calk^{n}_{+},$ $A\in\br^{n\times n}$ with
\begin{align*}
A_{ii}& =\sum_{j=1}^n |B_{ij}|-\mathcal{Y}_i,\; A_{ij}=-|B_{ij}|,\;\\
w & =\stack(k_i),\; b_i=-Q_i,
\end{align*}
can be rewritten as \eqref{stestaequ}. If we make the reasonable assumption that $\alpha_i<0$ for at least one node, $A$ satisfies Assumption~\ref{ass.1}. 
The reactive power balance \eqref{reacpowbal} has been recently employed in 
\cite{simpson16} to study long-term voltage stability.

We bring to the readers attention the fact that the coefficients $-b_i$ are the constant reactive powers extracted or injected into the network, being positive (capacitive) in the former case, and negative (inductive) in the latter. As indicated in Section~\ref{sec2} sharper results---{\em i.e.}, uniqueness of the equilibrium $\bar x_{\tt max}$, and a simpler structure of the set $\cale$ of initial conditions for the characteristic solutions---are available if the signs of the coefficients $b_i$ are known. Hence, the proposed conditions have a direct interpretation in terms of reactive power demand. 

Another observation is that the solution $\bar x_{\tt max}$ for the system \eqref{reacpowbal} represents the physically admissible steady state for the network with the highest values of voltage magnitudes at each node, which is the usually desired high-voltage operating point.

\subsection{Multi-terminal HVDC transmission networks with constant power devices}
\lab{subsec32}
%
An MT-HVDC network with $n$ power-controlled nodes ($\mathcal{P}$-nodes) and $s$ voltage-controlled nodes ($\mathcal{V}$-nodes), interconnected by $m$ RL transmission lines, can be modeled by \cite{sanchez18}:
\begin{equation}
\begin{split}
\tau \dot I_t&=-I_t- h(V),\\
L \dot I&=-R I+\mathcal B_{\mathcal{P}}^\top V+\mathcal B_{\mathcal{V}}^\top V_{\cal V},\\
C\dot V&=I_t-\mathcal B_{\cal P} I -GV,
\end{split} 
\label{mthvdc}
\end{equation}
where  $I\in\br^n$, $V\in \calk^n_{+}$, $I\in\br^m$ and $V_{\cal V}\in\mathbb{R}^{s}$. Also, the matrices $R$, $L$, $G$, $C$, and $\tau$ are diagonal, positive definite of appropriate sizes. The physical meaning of each state variable and of every matrix of parameters is given in Table~\ref{tab:state & param hvdc}.  Furthermore,  $\mathcal B=\stack\left(\cal B_{\cal V},\cal B_{\cal P} \right)\in\br^{(s+n)\times m}$ denotes the, appropriately split, node-edge incidence matrix of the network. The {\em open-loop} current injection at the power terminals is described by
\begin{equation*}
h(V)=\stack\left(\frac{P_i}{V_i} \right),
\end{equation*}
where $P_i\in\mathbb{R}$ denotes the power setpoint.\footnote{The first equation in \eqref{mthvdc} represents the simplified converter dynamics, see \cite[Section II, equation (18)]{sanchez18} and \cite[Figure 4]{sanchez18}. The converter usually has a PI current control, see the equations (27) and (28) of \cite{sanchez18}. For simplicity, we chose to study equilibria of the network without the PI. Nonetheless, our methodology applies also to the closed-loop scenario.}

\begin{table}\caption{Nomenclature for the model \eqref{mthvdc}.}
	\label{tab:state & param hvdc}
	\begin{tabular}{cc}
		\toprule
		&	State variables\\
		\bottomrule
		$I_t$ & $\mathcal{P}$-nodes injected currents\\
		$V$ & $\mathcal{P}$-nodes voltages\\
		$I$ & Line currents\\
		\toprule
		&	Parameters\\
		\bottomrule
		$L$ & Line inductances \\
		$C$ & $\mathcal{P}$-nodes shunt capacitances\\
		$R$ & Line resistances\\
		$G$ & $\mathcal{P}$-nodes shunt conductances\\
		$\tau$ & Converter time constants\\
		$V_{\cal V}$ & $\mathcal{V}$-nodes voltages\\
		\bottomrule
	\end{tabular}
	\vspace{-0.3cm}
	\centering
\end{table}

As done for the model \eqref{eq: dyn mod microgrid},  it can be shown by simple calculations that \eqref{mthvdc} admits an equilibrium  if and only if the system
\begin{equation}\label{eq:red 2}
0_n=-h( \bar V)-\left(\mathcal{B}_{\mathcal{P}}R^{-1}\mathcal{B}_{\mathcal{P}}^\top + G\right)\bar V-\mathcal{B}_{\mathcal{P}}R^{-1}\mathcal{B}_{\mathcal{V}}^\top V_{\mathcal{V}},
\end{equation}
has real solutions for $\bar V\in \calk^n_{+}$. Notice that \eqref{eq:red 2}  is equivalent  to the right hand side of \eqref{eq.1} if we define
\begin{equation*}
\begin{aligned}
x & := \bar{V},~ & A  &:= \mathcal{B}_{\mathcal{P}}R^{-1}\mathcal{B}_{\mathcal{P}}^\top + G,\\
b_i & := P_i,~ & w &:= -\mathcal{B}_{\mathcal{P}}R^{-1}\mathcal{B}_{\mathcal{V}}^\top V_{\mathcal{V}}.
\end{aligned}
\end{equation*}
Note that $\mathcal{B}_{\mathcal{P}}$ is an incidence matrix and $R$ and $G$ are diagonal positive definite matrices. Hence, the term $\mathcal{B}_{\mathcal{P}}R^{-1}\mathcal{B}_{\mathcal{P}}^{\top}$ is a Laplacian matrix and thus it is positive semidefinite.
Consequently, $A=A^\top$ is positive definite. Hence, Assumption \ref{ass.1} is satisfied and the results of Section \ref{sec2} can be used to analyze the existence of equilibria of the dynamical system \eqref{mthvdc}. This, through the computation of the solutions of $\dot x=f(x)$, taking $f$ as the right hand side of \eqref{eq:red 2}. 

In this scenario, the coefficients $-b_i$ are the powers extracted or injected into the network, being negative in the former case and positive in the latter. 

\subsection{DC microgrids with constant power loads}
\lab{subsec31}
%
A standard Kron-reduced model of a DC microgrid, with $n\geq 1$ converter-based distributed generation units, interconnected by $m\geq 1$ RL transmission lines, can be written as \cite{cucuzzella18}
\begin{equation}\label{eq: dyn mod microgrid}
\begin{aligned}
L_t\dot{I}_t & =-R_tI_t-V+u,\\
C_t\dot V & = I_t+\mathcal{B}I-I_{\text{ZIP}}(V),\\
L\dot{I} & =-\mathcal{B}^\top V-RI,
\end{aligned}
\end{equation}
where $I_t\in\br^n$, $V\in \calk^n_+,$ $u\in \calk^n_+$ and $I\in\br^m$ as well as $R_t$, $R$, $L_t$, $L$ and $C_t$ are diagonal, positive definite matrices of appropriate size. The physical meaning of each term appears in Table \ref{tab:state & param micro}. We denote by $\mathcal{B}\in\mathbb{R}^{n\times m}$, with $\mathcal{B}_{ij}\in\{-1,0,1 \}$,  the node-edge incidence matrix of the network. The load demand is described by a ZIP model, {\it i.e.,}
\begin{equation*}
I_{\text{ZIP}}(V)=\mathcal{Y}V+k+\stack\left(\tfrac{P_i}{V_i}\right),
\end{equation*}
where $\mathcal{Y}\in\mathbb{R}^{n\times n}$ is a diagonal positive semi-definite matrix, $k\in\mathbb{R}^{n}$ is a constant vector, and $P_i\in\mathbb{R}$.

\begin{table}\caption{Nomenclature for the model \eqref{eq: dyn mod microgrid}.}
	\label{tab:state & param micro}
\begin{tabular}{cc}
		\toprule
		&	State variables\\
		\bottomrule
$I_t$ & Generated currents\\
$V$ & Load and bus voltages\\
$I$ & Line currents\\
		\toprule
		&	Parameters\\
		\bottomrule
$L_t$ & Filter inductances \\
$L$ & Line inductances \\
$C$ & Shunt capacitances\\
$R_t$ & Filter resistances\\
$R$ & Line resistances\\
		\toprule
		&	External variables\\
		\bottomrule
$u$ & Control input (converter voltage)\\
$I_{\text{ZIP}}$ & \begin{tabular}{cc}
$\mathcal{Y}_i$: Constant impedance & $k_i$: Constant current\\
$P_i$: Constant power
\end{tabular}  \\
		\bottomrule
	\end{tabular}
		\vspace{-0.3cm}
	\centering
\end{table}

Some simple calculations show that, for a given $u=\bar u$ constant, the dynamical system \eqref{eq: dyn mod microgrid} admits a {real} steady state if and only if, the system 
\begin{equation}\label{eq: ss micro}
0_n=R_t^{-1}\left(\bar u -\bar V \right)-\mathcal{B}R^{-1}\mathcal{B}^\top \bar V-I_{\text{ZIP}}(\bar V),
\end{equation}
has real solutions for  $\bar V\in \calk^n_{+}$. Defining
\begin{equation*}
\begin{aligned}
x & :=\bar V\\
A & := R_t^{-1}+\mathcal{Y}+\mathcal{B}R^{-1}\mathcal{B}^\top\\
b_i & := P_i,\;i=1,\dots,n\\
w & := R_t^{-1}\bar u-k.
\end{aligned}
\end{equation*}
the system \eqref{eq: ss micro} can be written in the form \eqref{stestaequ}. Similarly as for the MT-HVDC model, it can be shown that $A$ is a positive definite matrix and, hence, satisfies the conditions in Assumption \ref{ass.1}.  Therefore, the results of Section \ref{sec2} can be applied to study the solutions of the steady-state equation \eqref{eq: ss micro}.

Once again, we underscore that the coefficients $-b_i$ are the active powers extracted or injected into the network, being negative in the former case, and positive in the latter.


\section{Numerical simulations}
\label{sec4}

In this section we present some numerical simulations that illustrate the results reported in Section \ref{sec2}.

\subsection{An RLC circuit with constant power loads}
\label{subsec41}
\begin{figure}
\begin{center}
\includegraphics[width=.66\linewidth]{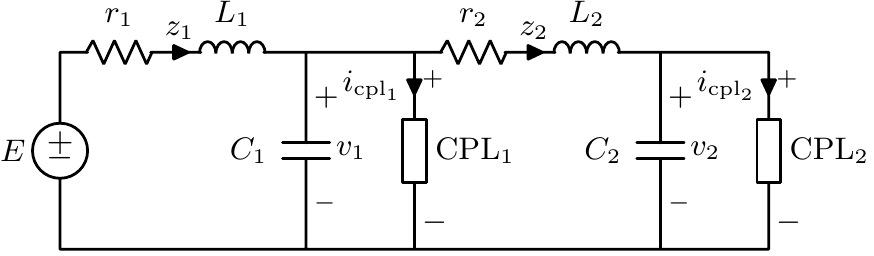}
	\caption{\normalsize DC Linear RLC circuit with two CPLs.}
	\label{fig: two port network}
\end{center}
\end{figure}

Consider the electrical network shown in Fig. \ref{fig: two port network}, which has been previously studied in \cite{barabanov16} as a benchmark example.  Its steady state is described by the system of quadratic equations
\begin{equation}\label{eq: thevenin 0}
\begin{aligned}
z & =-Yv+u\\
v_iz_i & =P_i > 0,\; i=1,2,
\end{aligned}
\end{equation}
where $z_i$ is the current of the  inductor $L_i$ and $v_i$ is the voltage of the capacitor $C_i$ and
\begin{align*}
Y  &=\left[
\begin{array}{cc}
 \frac{1}{r_2}+\frac{1}{r_1} & -\frac{1}{r_2} \\
 -\frac{1}{r_2} & \frac{1}{r_2}
\end{array}
\right],\;u  =\begmat{\tfrac{E}{r_1} \\0}.
\end{align*}
Defining
$$
x:=\begmat{ v_1 \\ v_2},\;A:=Y,\;b_i=P_i,\;w:=u,
$$
the algebraic equations \eqref{eq: thevenin 0} can be equivalently written in the form \eqref{stestaequ}.

\begin{figure}
\centering
\begin{subfigure}[b]{0.25\textwidth}
\includegraphics{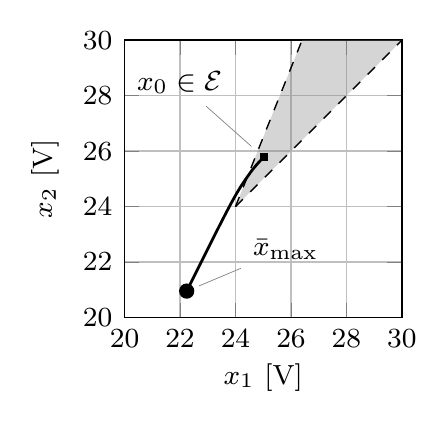}
\caption{}
\label{fig: ch ICs}
\end{subfigure}%
~
\begin{subfigure}[b]{0.25\textwidth}
\includegraphics{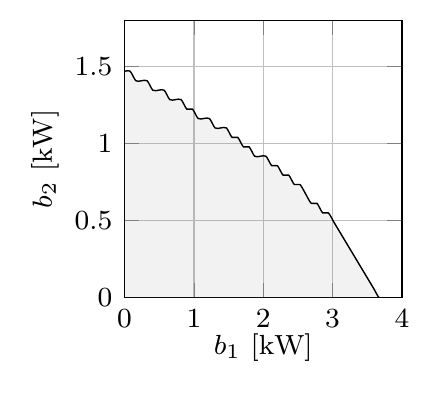}
\caption{}
\label{fig: ss two-port}
\end{subfigure}\\
\begin{subfigure}[b]{0.25\textwidth}
\includegraphics{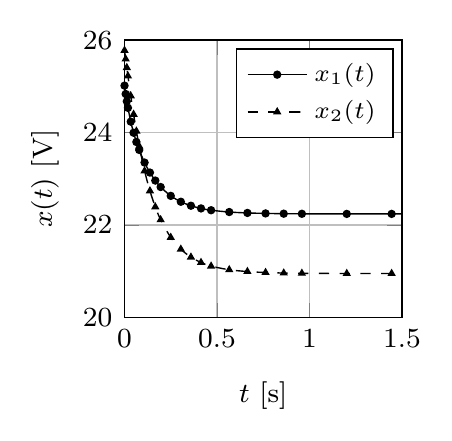}
\caption{}
\label{plot_exa1exp1plot1}
\end{subfigure}%
~
\begin{subfigure}[b]{0.25\textwidth}
\includegraphics{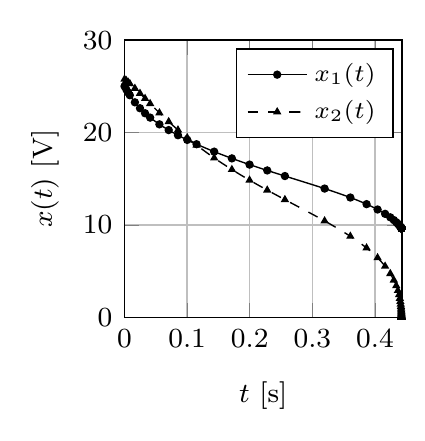}
\caption{}
\label{plot_exa1exp2plot1}
\end{subfigure}\\
\begin{subfigure}[b]{0.25\textwidth}
\includegraphics{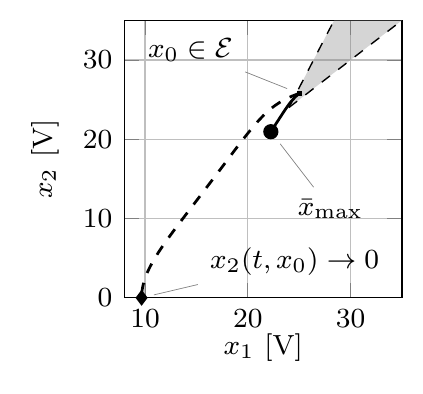}
\caption{}
\label{fig: phase-plot-both}
\end{subfigure}
{\scriptsize
\caption{Simulation results for the RLC circuit of Fig. \ref{fig: two port network}: (a) plot of a  a portion of the set $\mathcal{E}$ and a characteristic solution converging to $\bar{x}_{\tt max}$. (b)  Set of positive values (shadowed region) for $(b_1,b_2)$ for which the network admits an equilibrium. (c) Characteristic solution $x(t,x_0)$, with $b=(500,450)$, converging to the equilibrium point  $\bar x_{\tt max}$. (d) Characteristic solution $x(t,x_0)$, taking $b=(3000,1000)$,  with one of its components converging to zero in finite time, the system has no equilibrium points. (e) Phase-space plot of the characteristic solution $x(t,x_0)$ for two different values of $b$: one feasible and another one infeasible. Convergence to $\bar x_{\tt max}$ is observed in the former (solid curve), and convergence of the second component to zero is visualized in the latter (dashed curve).}}
\end{figure}

First, we compute the set $\cale$, given in \eqref{in.ch}, of initial conditions of the characteristic solutions as
{\footnotesize
$$
\cale=\left\{ x \in K_+^n\; |\;\left(\frac{1}{r_2}+\frac{1}{r_1}\right)x_1  -\frac{1}{r_2}x_2 >\frac{E}{r_1},\; -\frac{1}{r_2}x_1+ \frac{1}{r_2}x_2 >0\right\}. 
$$}
This set can also be written, in the simpler form
$$
\cale=\left\{ x \in K_+^n\; |\;E < x_1<x_2  <\frac{(r_1+r_2)}{r_1}x_1-\frac{r_2E}{r_1}\right\}. 
$$
A portion of this set, for the the values of the parameters given in Table \ref{tab:param2}, is shown in Fig. \ref{fig: ch ICs} together with a characteristic solution for $\dot x= f(x)$.\


Next, we verify numerically the procedure to test the existence of equilibria of the system \eqref{in.ch} suggested in Subsection \ref{subsec24}. Namely, taking an initial condition from the set $\cale$, we integrate the ODE to test whether one on the components of the state converges to zero in finite time, in which case there are no equilibria. On the other hand, if no component goes to zero, there are equilibria, and the trajectory will asymptotically converge to $\bar x_{\tt max}$. Notice that, according to Proposition \ref{th.main0}, since the coefficients $b_i>0$ this is the only equilibrium of the system.

\begin{table}\caption{Simulation Parameters of the multi-port network of Fig. \ref{fig: two port network}.}
	\label{tab:param2}
	\begin{tabular}{cccccc}
		\toprule
		$E$ (\SI{}{\volt}) & $r_1$ (\SI{}{\ohm})  & $L_1(\SI{}{\micro \henry})$  & $C_1(\SI{}{\milli \farad}) $   \\
		$24$ &$0.04$ & $78$ & $2$ \\
				\toprule
		 & $r_2$ (\SI{}{\ohm})  & $L_2(\SI{}{\micro \henry})$  & $C_2(\SI{}{\milli \farad}) $   \\
		 &$0.06$ & $98$ & $1$\\
		\bottomrule
	\end{tabular}
		\vspace{-0.3cm}
	\centering
\end{table}

Now, we recall that in \cite[Proposition 1 and 3]{barabanov16}, an LMI characterization for the existence of real solutions for \eqref{eq: thevenin 0} is given. Using this test, we obtain the set of (positive) values for $(b_1,b_2)$ for which there exists an equilibrium for the network---for any pair $(b_1,b_2)$ outside this region the equilibrium does not exist. The set of admissible powers is indicated by the shadowed region shown in Fig. \ref{fig: ss two-port}.


Next, we compute the solutions of the ODE \eqref{eq.1} in two scenarios. In the first case, we take $(b_1,b_2)=(500,450)$, which belongs to a feasible set according to Fig. \ref{fig: ss two-port}, then, the network has an equilibrium. We take the initial condition $x_0=(25.01,25.77) \in \cale$, and notice that none of the components of $x(t,x_0)$ approach zero---hence, we have the case {\bf III.(ii)} of Proposition \ref{th.main}, and $x(t,x_0)$ converges asymptotically to the unique equilibrium $\bar x_{\tt max}=(22.24,20.95)$, as shown in Fig \ref{plot_exa1exp1plot1}.


On the other hand, in Fig. \ref{plot_exa1exp2plot1},  we show the evolution of the same characteristic solution $x(t,x_0)$, but now taking $(b_1,b_2)=(3000,1000)$, which is outside the darkened region of the Fig. \ref{fig: ss two-port}, implying that the network admits no equilibria. Clearly, $x_2(t,x_0)$ converges to zero in finite time, as predicted by the case {\bf III.(i)}  of  Proposition \ref{th.main}.


Lastly, in Fig. \ref{fig: phase-plot-both} we present the plot of the characteristic solution $x(t,x_0)$ for the two scenarios just described, {\em i.e.,} with $(b_1,b_2)=(500,450)$, which is feasible, and with $(b_1,b_2)=(3000,1000)$ which is infeasible.


%
\subsection{An HVDC transmission system}

\begin{table}\caption{Numerical parameters associated with the edges for the network in Fig. \ref{fig:graph_example2}.}
	\label{tab:param_example2-2}
	\begin{tabular}{c|ccccc}
		\toprule
		Transmission line  & $e_1$  & $e_2$  & $e_3$ & $e_4$ & $e_5$  \\
		\bottomrule
  $r_i$ (\SI{}{\ohm})  & 0.9576 & 1.4365 & 1.9153 & 1.9153 & 0.9576
	\end{tabular}
		\vspace{-0.3cm}
	\centering
\end{table}

\begin{figure}
\begin{center}
\includegraphics[width=0.15\textwidth]{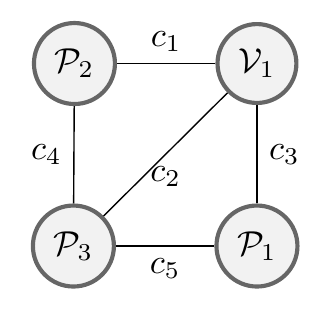}
\caption{Associated graph for the HVDC network studied in \cite[Section V]{sanchez18}.}
\label{fig:graph_example2}
\end{center}
\end{figure}

In this subsection we numerically evaluate the existence (and approximation) of equilibrium points for the particular HVDC system presented as an example in \cite[Fig. 5]{sanchez18}. The network, whose associated graph is shown in Fig. \ref{fig:graph_example2},  consists in four nodes $\mathcal{N}=\{\mathcal{V}_1,\mathcal{P}_1,\mathcal{P}_2,\mathcal{P}_3\}$, where $\mathcal{V}_1$ is a voltage controlled node with voltage $V_{\mathcal{V}}^{(1)}=E$, and $\mathcal{P}_1$, $\mathcal{P}_2$ and $\mathcal{P}_3$ are  power-controlled nodes with power $P_1$, $P_2$, and $P_3$, respectively. The network edges, representing the RL transmission lines, are $\mathbf{c}=\{c_1,c_2,...,c_5 \}$, with each $c_i$ having an associated pair of parameters $(r_i,L_i)$. If we assign arbitrary directions to the edges of the graph, then we can define an incidence matrix $\mathcal{B}=\stack\left(\mathcal{B}_{\mathcal{V}},\mathcal{B}_{\mathcal{P}} \right)$, where
\begin{align*}
\mathcal{B}_{\mathcal{V}} & =
\begin{bmatrix}
-1 & -1 & -1 & 0 & 0
\end{bmatrix},\\
\mathcal{B}_{\mathcal{P}} & = 
\begin{bmatrix}
0 & 0 & 1 & 0 & 1\\
1 & 0 & 0 & -1 & 0\\
0 & 1 & 0 & 1 & -1
\end{bmatrix}.
\end{align*}
Then, the elements of the algebraic system \eqref{eq:red 2}, which is codified by $f(x)=0$,  are given by
\begin{align*}
A &=\left[
\begin{array}{ccc}
 \gamma _1+\frac{1}{r_3}+\frac{1}{r_5} & 0 & -\frac{1}{r_5} \\
 0 & \gamma _2+\frac{1}{r_1}+\frac{1}{r_4} & -\frac{1}{r_4} \\
 -\frac{1}{r_5} & -\frac{1}{r_4} & \gamma _3+\frac{1}{r_2}+\frac{1}{r_4}+\frac{1}{r_5}
\end{array}
\right],\\
b &=\stack(P_i),~w = \stack\left(\frac{E}{r_3},\frac{E}{r_1},\frac{E}{r_2}\right),
\end{align*}
where $r_i$ and $\gamma_i$  are the diagonal elements of the matrices $R$ and $G$, respectively.

\begin{figure}
\begin{center}
\includegraphics[width=0.5\textwidth]{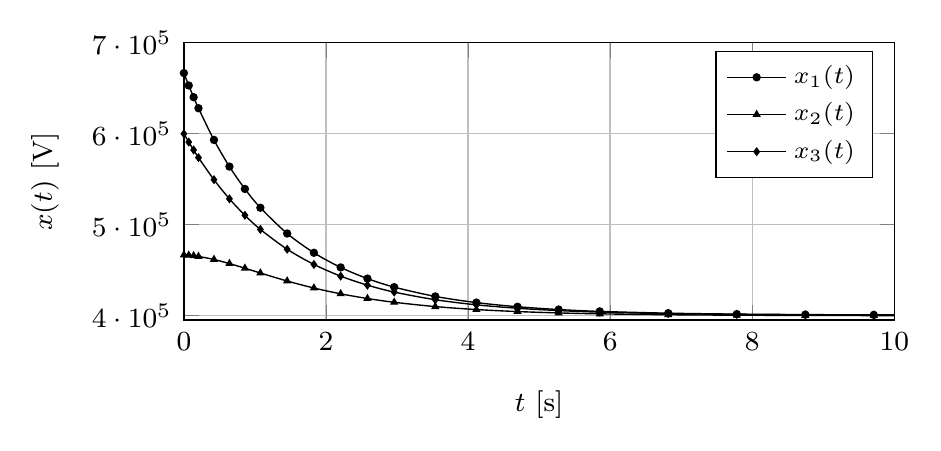}
\caption{Characteristic solution $x(t,x_0)$ converging to an equilibrium point. Once again, from Proposition \ref{th.main} we establish that $x(t,x_0)\rightarrow \bar{x}_{\tt max}$ as $t\rightarrow \infty$.}
\label{fig:example2plot1}
\end{center}
\end{figure}

Taking the numerical values shown in Tables \ref{tab:param_example2-1} and \ref{tab:param_example2-2}, we compute---through Lemma \ref{linear.lemma}---an initial condition $x_0\in\mathcal{E}$ given by
$$
x_0=10^{5}\cdot\stack(6.66, 4.66, 5.99).
$$
The particular solution $x(t,x_0)$ of $\dot x= f(x)$ is shown in Fig. \ref{fig:example2plot1}. Clearly, none of its components converges to zero. Then, by Proposition \ref{th.main}, we establish that the limit of this solution is the dominant equilibrium point, $\bar x_{\tt max}$, of the system. Its value is given by
$$
\bar{x}_{\tt max}=10^{5}\cdot \stack(4.0054, 3.9991, 4.0043).
$$

%

\begin{table}\caption{Numerical parameters associated with the nodes for the network in Fig. \ref{fig:graph_example2}.}
	\label{tab:param_example2-1}
	\begin{tabular}{c|ccccc}
		\toprule
		Power converter  & $\mathcal{V}_1$  & $\mathcal{P}_1$  & $\mathcal{P}_2$ & $\mathcal{P}_3$   \\
		\bottomrule
  $V_{\mathcal{V}}^{(i)}$ (\SI{}{\kilo \volt})  & 400 & - & - & -\\
  $P_i$ (\SI{}{\mega\watt})  & - &  -160 &140 & -180\\
    $\gamma_i$ (\SI{}{\micro\siemens})  & - & 0.02290 & 0.02290 & 0.3435
	\end{tabular}
		\vspace{-0.3cm}
	\centering
\end{table}


\label{subsec42}
%
\section{Conclusions}
\lab{sec5}
%
We have shown in the paper that the steady-state equations of several conventional and emerging power systems architectures satisfy a set of {\em nonlinear} algebraic constraints with a particular structure, denoted in the manuscript by $f(x)=0$. It was established that the associated ODE $\dot{x}=f(x)$  is a monotone dynamical system, for which we have described all possible scenarios for existence, uniqueness and stability of its equilibria. It was proven that if equilibria exist, then, there is a distinguished one, denoted by $\bar{x}_{\tt max}$, which dominates---component-wise---all the other ones and attracts all the ODE trajectories starting from a well-defined domain. We have further provided an algorithm to establish whether solutions of the ODE will converge to $\bar{x}_{\tt max}$ or not. 
By using the above-mentioned motivating correspondence, we have shown that if $x$ represents the voltage magnitudes in an AC or (HV)DC power system, then  $\bar{x}_{\tt max}$ corresponds to its unique long-term stable voltage equilibrium.

Finally, we have demonstrated via supporting numerical experiments on two benchmark power system models that our methodology performs very satisfactorily for realistic power system parametrizations. 

\appendices

%

\section{Proof of Lemma~\ref{lem.singular}}
\lab{appa}
The set
$
\Xi := \left\{ z \in \br^n: \det \left[ A - \diag\left(z_i\right)\right] =0 \right\}
$
is clearly closed and for any $i$ and given $z_j$'s with $j \neq i$, its section $\{z_i \in \br : \stack\left(z_1, \ldots,z_n\right) \in \Xi\}$ has no more than $n$ elements. So the Lebesgue measure of $\Xi$ is zero by the Fubini theorem. The function $x \in \mathcal{K}_+^n \mapsto g(x) := \stack \left( b_i x_i^{-2}\right)$ diffeomorphically maps $\mathcal{K}_+^n$ onto an open subset of $\br^n$. Hence the inverse image $\Xi_{\downarrow}:=g^{-1}(\Xi)$ is closed, has the
zero Lebesgue measure and, due to these two properties, is nowhere dense.
\par
Let $C$ be the set of all critical points of the semi-algebraic map \cite{Coste02} $x \in \mathcal{K}_+^n \mapsto h(x) := Ax + \stack \left( b_i x_i^{-2}\right) \in \br^n$, {\em i.e.}, points $x$ such that the Jacobian matrix $h^\prime(x)$ is singular. By the extended Sard theorem \cite{Kuorsi00}, the set of critical values $h(C)$ has the zero Lebesgue measure and is nowhere dense. Meanwhile, the restriction $h_{\calk_+^n \setminus C}$ is a local diffeomorphism and so the image $h(\Xi_{\downarrow} \setminus C)$ is nowhere dense and has the zero Lebesgue measure. It remains to note that the set of $w$'s for which Assumption~{\rm \ref{ass.zero}} does not hold lies in $h(\Xi_{\downarrow} \setminus C) \cup h(C)$. \qed

\section{Proof of Lemma \ref{linear.lemma}}
\lab{appd}
 The following system of linear inequalities is feasible
\begin{equation}
\label{lim.ineq}
A z >0, \qquad z>0 .
\end{equation}

{\bf Proof:} Suppose that the system \eqref{lim.ineq} is infeasible. Then two open convex cones $AK_+^n$ and $K_+^n$ are disjoint and so can be  separated by a hyperplane: there exists
\begin{equation}
\label{nonzero}
\tau \in \br^n, \qquad \tau \neq 0
\end{equation}
such that
$$
\tau^{\top} x \geq 0 \quad \forall x \in K_+^n, \qquad
\tau^{\top} x \leq 0 \quad \forall x \in AK_+^n.
$$
By continuity argument, these inequalities extend on the closures of the concerned sets:
\begin{align*}
 \tau^{\top} x \geq 0 & \quad  \forall x \in \overline{K}_+^n  = \{x : x_i \geq 0\},\\ \qquad \tau^{\top} x \leq 0 & \quad \forall x \in \overline{AK_+^n} \supset A \overline{K}_+^n.
\end{align*}
Here the first relation implies that $\tau \in \overline{K}_+^n$ and so $A \tau \leq 0$ by the second one. Hence $\tau\trs A \tau \leq 0$. Since $A$ is positively definite by Assumption~\ref{ass.1}, the last inequality yields that $\tau =0$, in violation of the second relation from \eqref{nonzero}. This contradiction completes the proof. \qed
\par
Based on any solution $z$ of \eqref{lim.ineq}, a solution of \eqref{in.ch} can be built in the form $x:= \mu z$ by picking $\mu>0$ so that
for all $i$,
\begin{gather*}
\mu (Az)_i > \langle w_i \rangle +\frac{\langle - b_i \rangle}{\mu z_i} \Leftrightarrow \mu^2 (Az)_i - \mu \langle w_i \rangle -\frac{\langle - b_i \rangle}{z_i} > 0
\\
\Leftrightarrow \mu > \frac{\langle w_i \rangle + \sqrt{\langle w_i \rangle^2+ 4 (Az)_i \frac{\langle - b_i \rangle}{z_i}}}{2 (Az)_i}.
\end{gather*}
Clearly, in the case $b_j>0$ for some $j$, then, the above inequality can be simplified as
$$
\mu > \frac{\langle w_i \rangle}{(Az)_i},~i=j.
$$
This concludes the proof.

\section{Technical facts needed to prove Propositions~\ref{th.main0} and \ref{th.main}}
\label{sec.proof}

In this section, we consider a $C^1$-map $g:\calk_+^n \to \br^n$ and provide a general study of the ODE
\begin{equation}
\label{eq.f}
\dot{x} = g(x), \qquad x \in \calk_+^n ,
\end{equation}
under  the following.
\begin{assumption}
\label{ass.tech1}
For any $x \in \calk_+^n$, the off-diagonal elements of the Jacobian matrix $\nabla g(x)$ are nonnegative.
\end{assumption}
\begin{assumption}
\label{ass.tech2}
For any $x \in \calk_+^n$, the Jacobian matrix $\nabla g (x)$ is symmetric.
\end{assumption}
For the convenience of the reader, we first recall several facts that will be instrumental in our study. The first group of them reflects that
the system \eqref{eq.f} is {\em monotone} (see \cite{SMI} for a definition).
\begin{proposition}
\label{prop.1}
Let Assumption~{\rm \ref{ass.tech1}} hold and let the order $\succ$ in $\br^n$ be either $\geq$ of $>$.
For any solutions $x_1(t), x_2(t), x(t)$ of \eqref{eq.f} defined on $[0,\tau], \tau>0$, the following implications hold
\begin{gather}
\label{nstr.ineq}
x_2(0) \succ x_1(0) \Rightarrow x_2(t) \succ x_1(t)\; \forall t \in [0,\tau],
\end{gather}
\begin{equation}
\label{dec.inder}
\begin{aligned}
\dot{x}(0) \prec 0 & \Rightarrow \dot{x}(t) \prec 0 \; \forall t \in [0,\tau],\\
 \dot{x}(0) \succ 0 & \Rightarrow \dot{x}(t) \succ 0 \; \forall t \in [0,\tau] ;
\end{aligned}
\end{equation}
\begin{equation}
\label{pos.invvrr}
\begin{aligned}
(x_+ >0~\wedge~ \varsigma & = \pm 1~\wedge~\varsigma g(x_+) >0 ) \\
 & \Rightarrow  \\
 \text{\rm the domain}  \; \Upsilon_\varsigma & :=\{x: \varsigma (x -  x_+) \succ 0 \}\cap \calk_+^n \;\\
&  \text{\rm is positively invariant}.
\end{aligned}
\end{equation}
\end{proposition}
{\bf Proof:} Relation \eqref{nstr.ineq} is given by Proposition~1.1 and Remark~1.1 in Chapter~3 of \cite{SMI}, whereas \eqref{dec.inder} is due to \cite[Prop.~2.1, Ch.~3]{SMI}. When proving \eqref{dec.inder}, we focus on $\varsigma=1$; the case $\varsigma =-1$ is treated likewise.
Let $x_\dagger(t), t \in [0,\theta)$ stand for the maximal solution of \eqref{eq.f} starting from $x_\dagger(0)=x_+$. Since $\dot{x}_\dagger(0) = g(x_+) \succ 0$, \eqref{dec.inder} guarantees that $x_\dagger(\cdot)$ constantly increases $\dot{x}_\dagger(t) \succ 0 \; \forall t \in [0,\theta)$ and so $x_\dagger(t) \succ  x_+ \; \forall t \in (0,\theta)$. Now let a solution $x(t), t \in [0,\tau], \tau \in (0,\infty)$ start in $\Upsilon_\varsigma$. Then $x(0) \succ x_\dagger(0)$ and $x(t) \succ x_\dagger(t) \succ x_+$ by \eqref{nstr.ineq}. So $x(t) \in \Upsilon_\varsigma$ for any $t \in [0,\tau] \cap [0,\theta)$. It suffices to show that $\tau < \theta$ if $\theta < \infty$.
\par
Suppose to the contrary that $\tau \geq \theta$.
Letting $t \to \theta-$, we see that $\|x(t)\| \to \infty$
by \cite[Th.~3.1, Ch.~II]{Hart82} since $x_\dagger(t) \succ x_+ >0$, and so $x(t) \succ x_\dagger(t) \Rightarrow \|x(t)\| \to \infty$. However, $\|x(t)\| \to \|x(\tau)\| < \infty$. This contradiction completes the proof.
\qed
\par
Let
$x(t,a), t \in [0,\tau_a)$ stand for the maximal solution of \eqref{eq.f} that starts at $t=0$ with $a>0$.
The distance $\inf_{x^\p \in A} \|x-x^\p\|$ from point $x \in \br^n$ to a set $A \subset \br^n$ is denoted by $\mathbf{dist}(x,A)$
\begin{corollary}
\label{corr.ineq}
Whenever $0 < a_1 \leq a \leq a_2$, we have $\tau_a \geq \min\{\tau_{a_1}, \tau_{a_2}\}$.
\end{corollary}
Claims similar to the following lemma can be inferred from the equivalences $\text{G}_{20}$ and $\text{I}_{27}$ in \cite[Th. 2.3, Ch. VI]{berman94} and (1.1) $\Leftrightarrow$ (1.2) in \cite[Prop.~1]{rantzer15}.
\begin{lemma}
A nonsingular matrix $A=A^\top$ with nonnegative off-diagonal elements is Hurwitz if
\begin{equation}
\label{fr.perron}
Ah > 0 \Rightarrow h \leq 0.
\end{equation}
\label{lem.frob}
\end{lemma}
{\bf Proof:} Since $A=A^\top$, the $A$-associated graph $\Gamma$ is undirected. (In $\Gamma$, the set of nodes is $1,\ldots,n$ and nodes $i \neq j$ are linked if and only if $a_{ij}=a_{ji} \neq 0$.) A proper permutation of the indices shapes all connected components of $\Gamma$ into intervals of the set of integers. Concurrently, the matrix $A$ takes a block diagonal form $A = \diag (A_1, \ldots, A_k)$
with irreducible blocks $A_s$.
\par
Now we pick $\alpha_i$ so large that all entries of $A_i^+:=A_i+\alpha_iI$ are nonnegative. The Perron-Frobenius theorem \cite[Th.~8.4.4]{Horn13} guarantees that $A^+_i$ has an eigenvalue $\lambda^+_i$ that dominates $\lambda^+_i > \lambda$ all other eigenvalues $\lambda$, and there is a $\lambda_i^+$-associated eigenvector $h_i >0$.
It follows that $\lambda_i:= \lambda^+_i - \alpha_i$ is a dominant eigenvalue of $A_i$ with the eigenvector $h_i$. Since $A$ is non-singular, so are $A_i$'s; hence $\lambda_i \neq 0$.
\par
For $h:= \stack (\{\mathbf{sgn} \lambda_j \times h_j\}_{j=1}^k)$, we have $Ah = \stack (\{|\lambda_j| \times h_j\}_{j=1}^k) >0$. So \eqref{fr.perron} yields that $\{\mathbf{sgn} \lambda_j \times h_j \leq 0 \Rightarrow  \mathbf{sgn} \lambda_j  =-1 \Leftrightarrow \lambda_j <0 \; \forall j$. Since the eigenvalue $\lambda_j$ is dominant, all eigenvalues of any block $A_j$ are negative. \qed
\begin{lemma}
\label{lem.locst}
Let Assumptions~{\rm \ref{ass.tech1}} and {\rm \ref{ass.tech2}} hold. Suppose that a solution $x(t), t \in [0,\infty)$ of \eqref{eq.f} decays $\dot{x}(t) <0 \; \forall t$ and converges to $\bar x >0$ as $t \to \infty$. Then $\bar x$ is an equilibrium of the ODE \eqref{eq.f}. If this equilibrium is hyperbolic, it is locally asymptotically stable.
\end{lemma}
{\bf Proof:} The first claim is given by \cite[Prop.~2.1, Ch.~3]{SMI}. By Lemma~\ref{lem.frob}, it suffices to show that $A:= \nabla g(\bar x)$ meets \eqref{fr.perron} to prove the second claim. Suppose to the contrary that there exists $h \in \br^n$ such that $Ah>0$ and $h_i >0$ for some $i$.  For $x_\ve^0 := \bar x + \ve h$ and small enough $\ve >0$, we have $g(x_\ve^0) = g(\bar x) + \ve A h + \text{\tiny $\mathcal{O}$}(\ve) = \ve A h + \text{\tiny $\mathcal{O}$}(\ve) >0, x^0_\ve >0$, and $x_{\ve,i}^0 > \bar x_{i}, x(0) \in \Upsilon_+ = \{ x: x> x_\ve^0 \}$. Since the set $\Upsilon_+$ is positively invariant by \eqref{pos.invvrr}, we infer that $x(t) \in \Upsilon_+ \Rightarrow x_i(t) > x_{\ve,i} >\bar x_{i}$, in violation of $x(t) \to \bar x$ as $t \to \infty$. This contradiction completes the proof. \qed
\par
For any $x^\prime \leq x^{\pp} \in \br^n$, we denote $\upharpoonright x^\prime , x^{\pp} \upharpoonleft := \{ x \in \br^n: x^\prime \leq x \leq x^{\pp} \}$.
\begin{lemma}
\label{lem.lse}
Suppose that Assumption~{\rm \ref{ass.tech1}} holds and $\bar x >0$ is a locally asymptotically stable equilibrium. Its domain of attraction $\mathscr{A}(\bar x) \subset \calk_+^n$ is open and
\begin{equation}
\label{inter,eq}
a_1,a_2 \in \mathscr{A}(\bar x) \wedge a_1 \leq a_2 \Rightarrow \upharpoonright a_1 , a_2 \upharpoonleft \subset \mathscr{A}(\bar x).
\end{equation}
\end{lemma}
{\bf Proof:}  Let $B(r,x)$ stand for the open ball with a radius of $r>0$ centered at $x$.
\par
For any $a \in \mathscr{A}(\bar x)$, we have $\tau_a = \infty$ and $x(t,a) \to \bar x$ as $t \to \infty$, whereas $B(2\ve,\bar x) \subset \mathscr{A}(\bar x)$ for a sufficiently small $\ve>0$ thanks to local stability of $\bar x$.
Hence there is $\theta>0$ such that $x(\theta,a) \in B(\ve,\bar x)$. By \cite[Th.~2.1, Ch.~V]{Hart82},
there exists $\delta>0$ such that whenever $\|a^\prime - a\| < \delta$, the solution $x(\cdot,a^\prime)$ is defined at least on $[0,\theta]$ and $\|x(\theta,a^\prime) - x(\theta,a)\| < \ve$. It follows that $x(\theta,a^\prime) \in B(2\ve,\bar x)$ and so $x(\cdot,a^\prime)$ is in fact defined on $[0,\infty)$ and converges to $\bar x$ as $t \to \infty$. Thus we see that $\|a^\prime - a\| < \delta \Rightarrow a^\prime \in \mathscr{A}(\bar x)$, {\em i.e.}, the set $\mathscr{A}(\bar x)$ is open.
\par
Let $a \in \upharpoonright a_1 , a_2 \upharpoonleft$. By Corollary~\ref{corr.ineq} and \eqref{nstr.ineq}, $\tau_a = \infty$ and
$x(t,a_1) \leq x(t,a) \leq x(t,a_2)\; \forall t \geq 0$. Letting $t \to \infty$ shows that $x(t,a) \to \bar x$ and so $a \in \mathscr{A}(\bar x)$.
\qed
\begin{lemma}
\label{lem.retr} Let $\ov{x}_1 \leq \ov{x}_2$ and let $D\subset \Xi :=
\upharpoonright \ov{x}_1, \ov{x}_2 \upharpoonleft$ be an open (in $\Xi$) set
such that {\bf (i)} $\upharpoonright x^\prime, x^{\pp} \upharpoonleft
\subset D \;\forall x^\prime, x^{\pp} \in D$; {\bf (ii)} either $\ov{x}_{1}
\in D$ or $\ov{x}_{2} \in D$; {\bf (iii)} $D \neq \Xi$. Then there
exists a continuous map $M: \Xi \to \Xi$ such that $ M\left[\Xi\right]
\subset \Xi_- := \Xi \setminus D$ and $M[x] = x \quad \forall x \in \Xi_-.
$\footnote{In brief, this lemma says that $\Xi_-$ is a retract of the convex set $\Xi$.}
\end{lemma}
{\bf Proof: } Let $\ov{x}_2 \in D$ for the definiteness; then $\ov{x}_1 \not\in D$ by (i)
and (iii). It can be evidently assumed that $0=\ov{x}_1 < \ov{x}_2$. We
denote $\chi_x(\theta):= \max\{ x-\theta\zeta; 0 \}$, where $\zeta :=\stack
(1,\ldots,1)$ and the $\max$ is meant component-wise. Evidently, $\Theta(x)
:= \big\{ \theta \geq 0 : \chi_x(\theta) \in D \big\} = [0,\tau(x)) \; x \in
D $, where $0< \tau(x) < \infty$. For $x\not\in D$, we put $\tau(x) := 0$.
We are going to show first that the function $\tau(\cdot)$ is continuous on
$\Xi$. To this end, it suffices to prove that $\tau(\bar x) = \tau_\ast$
whenever $$ \bar x = \lim_{k \to \infty} x_k, \quad x_k \in \Xi , \quad
\text{and} \quad \tau_\ast = \lim_{k \to \infty} \tau(x_k) . $$ Passing to a
subsequence ensures that either $x_k \not\in D \; \forall k$ or $x_k \in D
\; \forall k$. In the first case, $\bar x \not\in D$ since $D$ is open. Then
$\tau(\bar x) =0 = \tau(x_k) = \tau_\ast$. Let $ x_k \in D \; \forall k$.
Since $\chi_{x_k}[\tau(x_k)]\not\in D$ and $D$ is open, letting $k \to
\infty$ yields $ \chi_{\bar x}[\tau_\ast] \not\in D \Rightarrow \tau(\bar x)
\leq \tau_\ast$. So the claim holds if $\tau_\ast =0$. If $\tau_\ast
>0$, we pick $0< \theta < \tau_\ast$.
Then $\theta < \tau(x_k)$ for $k \approx \infty$, {\em i.e.}, $\chi_{x_k}(\theta)
\in D$. Let $x_{\nu,i}$ be the $i$th component of $x_\nu \in \br^p$. Then
\begin{align*}
\tau^\prime_k & := \max \big\{\tau \geq 0 : \chi_{\bar x}(\tau)
\geq \chi_{x_k}(\theta)\big\}\\
& = \max_{i: x_{k,i} \geq \theta} \big[
x_{\ast,i} - x_{k,i} + \theta \big].
\end{align*}
Here the second $\max$ is over
a nonempty set since $\chi_{x_k}(\theta) \in D \not\ni 0$. Thus
$\tau_k^\prime \to \theta$ as $k \to \infty$. By (i),
$\chi_{\bar x}(\tau^\prime_k) \in D$ and so $\tau(\bar x) \geq \tau^\prime_k
\overset{k \to \infty}{=\!=\!=\!\Rightarrow} \tau(\bar x) \geq \theta \;
\forall \theta < \tau_\ast \Rightarrow \tau(\bar x) \geq \tau_\ast
\Rightarrow \tau(\bar x) = \tau_\ast$. Thus the function $\tau(\cdot)$ is
continuous indeed. The needed map $M$ is given by $M(x):=
\chi_x[\tau(x)]$.\qed
\begin{lemma}
\label{lem.repe1} Let Assumption~{\rm \ref{ass.tech1}} hold and $0< \ov{x}_1 \leq \ov{x}_2, \ov{x}_1 \neq
\ov{x}_2$ be two locally asymptotically stable equilibria. Then there exists a third equilibrium
$\bar x$ in between them $\ov{x}_1 \leq \bar x \leq \ov{x}_2, \bar x \neq
\ov{x}_1, \ov{x}_2$.
\end{lemma}
{\bf Proof:} By Lemma~\ref{lem.lse}, the set $D_i:=\mathscr{A}(\ov{x}_i)
\cap \Xi, i=1,2$ meets the assumptions of
Lemma~\ref{lem.retr}, which associates this set with a map $M_i$. Since the sets $D_i$ are open and
disjoint, they do not cover the connected set $\Xi$. So the set $\Xi_{\blacklozenge}:=\Xi \setminus (D_1\cup D_2)$ of all fixed points of the map $M = M_1\circ M_2$ is non-empty and compact.
\par
For all $a \in \Xi$,
the solution $x(\cdot,a)$ is defined on $[0,\infty)$ by Corollary~\ref{corr.ineq} and $x(t,a) \in \Xi$ by \eqref{nstr.ineq}. So the flow $\{\Phi_t(a) := x(t,a)\}_{t \geq 0}$ is well defined on $\Xi$, acts from $\Xi$ into $\Xi$, and is continuous by \cite[Th.~2.1, Ch.~V]{Hart82}. The sets $D_i$ are positively and negatively invariant with respect to it:
\begin{align*}
a \in & D_i \Rightarrow \Phi_t(a) \in D_i \; \forall  t \geq 0,\\
 a \in & \Xi \wedge \big[ \exists t \geq 0: \Phi_t(a) \in D_i \big] \Rightarrow a \in D_i.
\end{align*}
It follows that $\Xi_{\blacklozenge}$ is positively invariant with respect to this flow.
By the Brouwer fixed point theorem, the continuous map $\Phi_t\circ M : \Xi \to \Xi_{\blacklozenge} \subset \Xi
$ has a fixed point $a_t= \Phi_t\circ M (a_t) \in \Xi$.
Since $M(a_t) \in \Xi_{\blacklozenge}$ and $\Phi_t(\Xi_{\blacklozenge}) \subset \Xi_{\blacklozenge}$, we see that $a_t \in \Xi_{\blacklozenge}$ and so $M(a_t) = a_t$ and $a_t = \Phi_t(a_t)$.
\par
Since $\Xi_{\blacklozenge}$ is compact, there exists a sequence $\{t_k >0\}_{k=1}^\infty$ such that $t_k \to 0$ and $a_{t_k} \to \bar x$ as $k \to \infty$ for some point $\bar x \in \Xi_{\blacklozenge}$. Since $\ov{x}_1, \ov{x}_2 \not\in \Xi_{\blacklozenge}$, we have $\bar x \neq \ov{x}_1, \ov{x}_2$; meanwhile $\bar x \in \Xi_{\blacklozenge} \subset \Xi \Rightarrow \ov{x}_1 \leq \bar x \leq \ov{x}_2$. Furthermore,
 \begin{align*}
0 & = t_k^{-1} \left[ \Phi_{t_k}(a_{t_k}) -  a_{t_k} \right]\\
&  = t_k^{-1} \int_0^{t_k} g [ x(t,a_{t_k})] \; dt \xrightarrow{k \to \infty} g(\bar x).
 \end{align*}
Thus we see that $g(\bar x)=0$, {\em i.e.}, $\bar x$ is an equilibrium. \qed

\section{Proofs of Propositions~\ref{th.main0} and \ref{th.main}}
\lab{appc}
Now we revert to study of the system \eqref{eq.1} under the Assumptions \ref{ass.1} and \ref{ass.2}.
\begin{lemma}
\label{decay.lemma}
Suppose that $y$ belongs to the set \eqref{in.ch}. There exists $\theta \in (0,1)$ such that the domain $\Xi_-(\theta) := \{x: 0 < x \leq \theta y\}$ is globally absorbing, {\em i.e.}, the following statements hold:
\begin{enumerate}[{\bf (i)}]
\item This domain is positively invariant: if a solution starts in $\Xi_-(\theta)$, it does not leave $\Xi_-(\theta)$;
\item Any solution defined on $[0,\infty)$ eventually enters $\Xi_-(\theta)$ and then never leaves this set.
\end{enumerate}
\end{lemma}
{\bf Proof:}
Thanks to \eqref{in.ch}, there exists $\delta >0$ such that
\begin{equation}
\label{in.ch1}
 Ay > \stack \left( \langle w_i \rangle +\frac{\langle -b_i \rangle}{y_i} + 3\delta\right).
\end{equation}
We also pick $\theta \in (0,1)$ so close to $1$ that
\begin{equation}
\label{close}
[\theta -1] \langle w_i \rangle_+ + \delta \theta \geq 0, \quad
[\theta -\theta^{-1}] \langle - b_i\rangle y_i^{-1} + \delta \theta \geq 0 \qquad \forall i.
\end{equation}
Let $x(\cdot)$ be a solution of \eqref{eq.1}. By the Danskin theorem \cite{Danskin66}, the function
$
\varrho(t) := \max_{i=1,\ldots,n} x_i(t)/y_i
$
is absolutely continuous and for almost all $t$, the following equation holds
\begin{equation}
\label{danskin}
\begin{aligned}
\dot{\varrho}(t) & = \max_{i\in I(t)} \dot{x}_i(t)/y_i, \quad \text{where}\\
  I(t) & := \left\{i : x_i(t)/y_i  = \varrho(t) \right\}.
\end{aligned}
\end{equation}
For any $i \in I(t)$ and $j$, we have $x_i(t) = y_i \varrho(t), x_j(t) \leq y_j \varrho(t)$, and
\begin{equation}
\label{ind.du}
\begin{aligned}
\dot{x}_i(t) \overset{\text{\eqref{eq.1}}}{=} & - a_{ii} x_i(t) + \sum_{j \neq i} \underbrace{[\hspace{10.0pt}-a_{i,j}\hspace{10.0pt}]}_{\geq 0 \,\text{by Asm.~\ref{ass.1}}} x_j(t) -\frac{b_i}{x_i(t)} + w_i\\
\leq & -  \varrho(t) \left(a_{ii} y_i +  \sum_{j \neq i} a_{i,j} y_j \right) - \varrho(t)^{-1}\frac{b_i}{y_i} + w_i\\
\overset{\text{\eqref{in.ch1}}}{\leq} & - \varrho(t) \left[\langle w_i \rangle + \frac{\langle - b_i\rangle}{y_i} +3\delta \right] + \varrho(t)^{-1}\frac{\langle - b_i\rangle}{y_i} +\langle w_i \rangle\\
= & -\delta \varrho(t)
- \left\{ [\varrho(t) -1] \langle w_i \rangle + \delta \varrho(t) \right\}+\cdots\\
& \cdots  - \left\{ [\varrho(t) -\varrho(t)^{-1}] \frac{\langle - b_i\rangle}{y_i} + \delta \varrho(t) \right\}.
\end{aligned}
\end{equation}
Hence whenever $\varrho(t) \geq \theta \in (0,1)$,
\begin{align*}
\dot{x}_i(t) & \leq -\delta \varrho(t)
- \left\{ [\theta -1] \langle w_i \rangle_+ + \delta \theta \right\}+\cdots\\
& \cdots - \left\{ [\theta -\theta^{-1}] \frac{\langle b_i\rangle_-}{y_i} + \delta \theta \right\} \overset{\text{\eqref{close}}}{\leq} - \delta \varrho(t).
\end{align*}
So by invoking \eqref{danskin}, we infer that
$
\varrho(t) > \theta \Rightarrow \dot{\varrho}(t) \leq - \delta \varrho(t) \leq - \delta \theta.\footnote{In fact, this implication holds for almost all $t$ such that the premises are true.}
$
Claims (i) and (ii) are immediate from this entailment. \qed
\begin{lemma}
\label{lem.repe2}
Claim {\bf II)} of Proposition~{\rm \ref{th.main}} holds.
\end{lemma}
{\bf Proof:} This is immediate from \eqref{dec.inder} since for any characteristic solution $x(\cdot)$ and $y:=x(0)$,
\begin{align*}
\dot{x}(0) \overset{\text{\eqref{eq.1}}}{=} & - A y + \stack\left( - \frac{b_i}{y_i} + w_i\right)\\
 \leq &
- A y + \stack\left( \frac{\langle - b_i \rangle}{y_i} + \langle w_i \rangle \right) \overset{\text{\eqref{in.ch}}}{<} 0. \qquad \Box
\end{align*}
\begin{lemma}
\label{lem.nonexst}
Suppose that a solution $x(\cdot)$ of \eqref{eq.1} is defined on $[0,\tau)$ with $\tau<\infty$ but cannot be extended to the right.
Then there is $i$ such that $b_i>0$ and $x_i(t) \to 0, \dot{x}_i(t) \to - \infty$ as $t \to \tau-$.
\end{lemma}
{\bf Proof:} By Lemma~\ref{linear.lemma}, there exists a solution $y>0$ of \eqref{in.ch}. Via multiplying $y$ by a large enough
factor, we ensure that $y>x(0)$. Let $x_{\uparrow}(\cdot)$ be the characteristic solution starting with $x_{\uparrow}(0) = y$. By Lemma~\ref{lem.repe2}, $x_{\uparrow}(t) \leq y$ for $t \geq 0$, whereas $x(t) \leq x_{\uparrow}(t)$ on the intersection of the domains of definitions of $x(\cdot)$ and $x_{\uparrow}$ by \eqref{nstr.ineq}. Then \cite[Th.~3.1, Ch.~II]{Hart82} ensures that $x(t)$ converges to the boundary of $\calk_+^n$ as $t \to \tau-$ and is bounded. In other words,
\begin{equation}
\label{tobaund}
\min_i x_i(t) \to 0 \quad \text{as} \quad t \to \tau- , \qquad c:= \sup_{t \in [0,\tau)} \|x(t)\| < \infty.
\end{equation}
\par
Meanwhile putting $W:= \max_i \big[|w_i|+ c\sum_j |a_{ij}| \big] $, we see that
\begin{align*}
\dot{x}_i(t) \overset{\text{\eqref{ind.du}}}{=} & - \sum_{j} a_{i,j} x_j(t) -\frac{b_i}{x_i(t)} + w_i\\
 \in & ~ \left[ - W -\frac{b_i}{x_i(t)}, W -\frac{b_i}{x_i(t)}\right] ,
\end{align*}
\begin{equation}
\label{dfd1}
\begin{aligned}
b_i<0 \wedge x_i(t) \leq \frac{|b_i|}{2W} \Rightarrow \dot{x}_i(t) \geq W > 0,
\end{aligned}
\end{equation}
\begin{equation}
\label{dfd2}
\begin{aligned}
b_i>0 \wedge x_i(t) \leq \frac{|b_i|}{2W} \Rightarrow \dot{x}_i(t) \leq - \frac{b_i}{2 x_i(t)} < 0\\
\Rightarrow x_i^2(\theta) \leq x_i^2(t) - b_i(\theta - t) \; \forall \theta \in [t,\tau).
\end{aligned}
\end{equation}
Here \eqref{dfd1} implies that $x_i(t)$ is separated from zero if $b_i <0$. Hence \eqref{tobaund} yields that there exists $i$ such that $b_i>0$ and for any $\ve>0$, arbitrarily small left vicinity $(\tau-\delta, \tau), \delta \approx 0$ of $\tau$ contains points $t$ with $x_i(t) < \ve$. Then for $\ve < \frac{|b_i|}{2W}$, formula \eqref{dfd2} guarantees that $x_i(t^\prime) < \ve \; \forall t^\prime \in (t, \tau)$. Overall, we see that $x_i(t) \to 0$ as $t \to \tau-$; then $\dot{x}_i(t) \to - \infty$ as $t \to \tau-$ by \eqref{dfd2}. \qed
\begin{lemma}
\label{lem.alsost}
{\bf (i)} Stable equilibria of \eqref{eq.1} (if exist) are locally asymptotically stable. {\bf (ii)}
 Let $0< x^{-} \leq x^{0} \leq x^{+}$ be equilibria of \eqref{eq.1}. If $x^{\pm}$ are stable and all $b_i$'s are of the same sign, $x^0$ is also stable.
\end{lemma}
\par
{\bf Proof:} By Assumption~\ref{ass.2} and \eqref{eq.1}, the Jacobian matrix
\begin{equation}
\label{jack}
\begin{aligned}
\nabla f(x) & = A(k):= -A + \mathbf{diag} \left[k_i\right],\\
 k & :=k(x):=  \stack \left( b_i x_i^{-2} \right)
\end{aligned}
\end{equation}
has no eigenvalues with the zero real part at any equilibrium $x$. So an equilibrium $x$ is locally stable if and only if the matrix \eqref{jack} is Hurwitz and so $x$ is locally asymptotically stable.
Meanwhile, $A\trs = A$ by Assumption~\ref{ass.1}. So this local stability, in turns, holds if and only if
the following quadratic form in $h \in \br^n$ is negatively definite
$$
Q_x(h) := - h\trs A h + \sum_{i=1}^n k_i(x) h_i^2.
$$
Thus both forms $Q_{x^\pm}$ are negatively definite. Meanwhile, $k_i(x^0) \leq k_i(x^-) \forall i$ if $b_i>0 \; \forall i$, whereas
$k_i(x^0) \leq k_i(x^+) \forall i$ if $b_i<0 \; \forall i$. In any case, $Q_{x^0}$ is upper estimated by a negatively definite quadratic form (either $Q_{x^-}$ or $Q_{x^+}$) and so is negatively definite as well. \qed
\begin{corollary}
\label{cor.single}
Suppose that $0< x^{(0)} \leq x^{(1)}$ are stable equilibria of \eqref{eq.1} and all $b_i$'s are of the same sign. Then $x^{(0)} = x^{(1)}$.
\end{corollary}
{\bf Proof:} Suppose to the contrary that $x^{(0)} \neq x^{(1)}$. By Lemma~\ref{lem.repe1} and (i) of Lemma~\ref{lem.alsost}, there exists one more equilibrium $x^{(1/2)}$ in between $x^{(0)}$ and $x^{(1)}$, {\em i.e.}, $x^{(0)} \leq x^{(1/2)} \leq x^{(1)}$ and $x^{(1/2)} \neq x^{(0)}, x^{(1)}$. By (ii) of Lemma~\ref{lem.alsost}, this newcoming equilibrium $x^{(1/2)}$ is stable. This permits us to repeat the foregoing arguments first for $x^{(0)}$ and $x^{(1/2)}$ and second for $x^{(1/2)}$ and $x^{(1)}$. As a result, we see that there exist two more stable equilibria $x^{(1/4)} \in \upharpoonright x^{(0)}, x^{(1/2)}\upharpoonleft$ and $x^{(3/4)} \in \upharpoonright x^{(1/2)}, x^{(1)}\upharpoonleft$ that differ from all previously introduced equilibria.
This permits us to repeat the foregoing arguments once more to show that there exist stable equilibria $x^{(1/8)}, x^{(3/8)}, x^{(5/8)}, x^{(7/8)}$ such that $x^{(i/8)} \leq x^{(j/8)}\; \forall 0 \leq i \leq j \leq 8$ and $x^{(i/8)} \neq x^{(j/8)}\; \forall 0 \leq i,j \leq 8, i \neq j$.
By continuing likewise, we assign a stable equilibrium $x^{(r)}$ to any number $r\in [0,1]$ whose representation in the base-2 numeral system is finite ({\em i.e.}, number representable in the form $r = j2^{-k}$ for some $k=1,2,\ldots$ and $j = 0,\ldots,2^k$) and ensure that these equilibria are pairwise distinct and depend on $r$ monotonically: $x^{(r)} \leq x^{(\varrho)}$ whenever $0 \leq r \leq \varrho \leq 1$.
\par
Since all they lie in the compact set $\upharpoonright x^{(0)}, x^{(1)}\upharpoonleft$, there exists a sequence $\{r_k\}_{k=1}^\infty$ of pairwise distinct numbers $r$'s for which $\exists \bar x = \lim_{k \to \infty} x^{(r_k)}$. Then $\bar x \in \upharpoonright x^{(0)}, x^{(1)}\upharpoonleft$ and so $\bar x>0$ and
$f(\bar x) = \lim_{k \to \infty} f[x^{(r_k)}] =0$, {\em i.e.}, $\bar x$ is an equilibrium. Then the Jacobian matrix $\nabla f (\bar x)$ is nonsingular, as was remarked just after \eqref{jack}. However, this implies that in a sufficiently small vicinity $V$ of $\bar x$, the equation $f(x)=0$ has no roots apart from $\bar x$ in violation of $x^{(r_k)} \in V \; \forall k \approx \infty$ and $x^{(r_k)} \neq x^{(r_l)}\; \forall k \neq l$. The contradiction obtained completes the proof. \qed
\par
{\bf Proof of Proposition \ref{th.main}:} {\bf Claim I)} is justified by Lemma~\ref{linear.lemma}.
\\
{\bf Claim II)} is justified by Lemma~\ref{lem.repe2}. By II), the limit $\bar x$ from \eqref{limt} exists and $\bar x \geq 0$. 
\\
{\bf Claim III)} Let $x(t), t \in [0,t_f)$ be a characteristic solution. If $t_f < \infty$, then III.i) of Proposition \ref{th.main} holds  by Lemma~\ref{lem.nonexst}. Suppose that $t_f = \infty$. Then the limit $\bar x$ from \eqref{limt} exists due to II) of Proposition \ref{th.main}, and $\bar x \geq 0$.
We are going to show that in fact $\bar x >0$.
\par
Suppose to the contrary that $\bar x_i = 0$ for some $i$. Then $x_i(t) \to 0$ as $t \to \infty$, \eqref{dfd1} means that $b_i>0$, and \eqref{dfd2} (where $\tau = \infty$ now) implies that $\|x(\theta)\|^2$ assumes negative values for large enough $\theta$. This assures that 
$\bar x >0$ and so \eqref{limt} does hold. By Lemma~\ref{lem.locst}, $\bar x$ is an equilibrium. 
\par
Now suppose that III.i) holds for a characteristic solution $x_\dagger(\cdot)$. 
Suppose that there is another characteristic solution $x(\cdot)$ for which III.i) is not true. Then $x(\cdot)$ is defined on $[0,\infty)$ by Lemma~\ref{lem.nonexst} and also $\exists \bar x = \lim_{t \to \infty} x(t) >0$ by the foregoing. By (ii) of Lemma~\ref{decay.lemma} (with $y:= x_\dagger(0)$), $x(\sigma) \leq \theta x_\dagger (0) \leq x_\dagger(0)$ for large enough $\sigma$. By applying \eqref{nstr.ineq} to $x_1(t):= x(t+\sigma)$ and $x_2(t) = x_\dagger(t)$, we see that $x(t+\sigma) \leq x_\dagger(t)$ and so $x_i(t)$ goes to zero in a finite time, in violation of $\bar x$. This contradiction proves that III.i) holds simultaneously for all characteristic solutions.
\par
Since III.i) and III.ii) are mutually exclusive and complementary, we see that either III.i) holds for all characteristic solutions, or 
III.ii) holds for all of them. 
\par
Finally, suppose that III.ii) holds. As was shown in the penultimate paragraph, $x(t+\sigma) \leq x_\dagger(t)$ for any two characteristic solutions $x(\cdot)$ and $x_\dagger(\cdot)$. Hence $\lim_{t \to \infty} x(t) \leq \lim_{t \to \infty} x_\dagger(t)$. By flipping $x(\cdot)$ and $x_\dagger(\cdot)$ here, we see that these limit coincide, {\it i.e.,} the limit \eqref{limt} is the same for all characteristic solutions.   
\\
{\bf Claim IV)} is straightforward from Lemmas~\ref{decay.lemma} and \ref{lem.nonexst} since any equilibrium is associated with a constant solution defined on $[0,\infty)$.
\\
{\bf Claim V)} Suppose that III.ii) holds. Let $\bar{x}_{\tt max}$ stand for the limit \eqref{limt}.
By (II) and Lemmas~\ref{lem.locst} and \ref{lem.alsost}, $\bar{x}_{\tt max}$ is a locally asymptotically stable equilibrium. 
Let us consider a solution $x(\cdot)$ defined on $[0,\infty)$ and a characteristic solution $x_\dagger(\cdot)$.
By retracing the above arguments based on (ii) of Lemma~\ref{decay.lemma}, we see that $x(\varsigma +t) \leq x_\dagger(t) \; \forall t \geq 0$ for some $\varsigma \geq 0$. By considering here a constant solution $x(\cdot)$ and letting $t \to \infty$, we see that $\bar x_{\max}$ dominates any other equilibrium.  
\par
Now suppose that $x(0) \geq \bar x_{\max}$. By \eqref{nstr.ineq}, $x(t) \geq \bar x_{\max}$ on the domain $\Delta$ of definition of $x(\cdot)$ and so $\Delta = [0, \infty)$ by Lemma~\ref{lem.nonexst}. Thus we see that $x_{\max} \leq x(\varsigma +t) \leq x_\dagger(t) \; \forall t \geq 0$ for some $\varsigma \geq 0$. It follows that $x(t) \to x_{\max}$ as $t \to \infty$, {\it i.e.,} the equilibrium $x_{\max}$ is attractive from the right by Definition~\ref{def.prpr}.
\par
It remains to show that there exist only finitely many equilibria $\bar x^k$. 
Suppose the contrary. Since all equilibria lie in the compact set $\{x: 0 \leq x \leq \bar x_{\max}\}$, there exists an infinite sequence $\{\bar x^{k_s}\}_{s=1}^\infty$ of pairwise different equilibria that converges $\bar x^{k_s} \to \bar x$ as $t \to \infty$ to a point $\bar x \geq 0$.  
The estimates \eqref{dfd1}, \eqref{dfd2} applied to any equilibrium solution $x(\cdot)$ assure that $x_i \geq |b_i|/(2W)$ on it, where $W:= \max_i \big[|w_i|+ c\sum_j |a_{ij}| \big] $ and $c$ is any upper bound on $\|x(t)\|$. For the solutions related to the convergent and so bounded sequence 
$\{\bar x^{k_s}\}_{s=1}^\infty$, this bound can be chosen common. As a result, we infer that $\bar x >0$ 
and so $f(\bar x) = \lim_{s \to \infty} f[x^{k_s}] =0$, {\em i.e.}, $\bar x$ is an equilibrium. Then the Jacobian matrix $\nabla f (\bar x)$ is nonsingular, as was remarked just after \eqref{jack}. This implies that in a sufficiently small vicinity $V$ of $\bar x$, the equation $f(x)=0$ has no roots apart from $\bar x$, in violation of $x^{k_s} \in V \; \forall s \approx \infty$ and $x^{k_s} \neq x^{k_r}\; \forall s \neq r$. This  contradiction completes the proof. \qed
\par
{\bf Proof of Proposition \ref{th.main0}:} This proposition is immediate from Proposition \ref{th.main}. \qed


%


%
\section*{Acknowledgment}
This paper is partly supported by the Ministry of Education and Science of Russian Federation (14.Z50.31.0031, goszadanie no. 8.8885.2017/8.9), NSFC (61473183, U1509211). {The work of Juan E. Machado was supported by the Mexican government through the National Council of Science and Technology (or CONACyT for its acronym in Spanish)}.
%
\bibliographystyle{IEEEtran} 
\bibliography{bib_matveev}


\begin{IEEEbiography}
[{\includegraphics[width=1in,height=1.25in,clip,keepaspectratio]{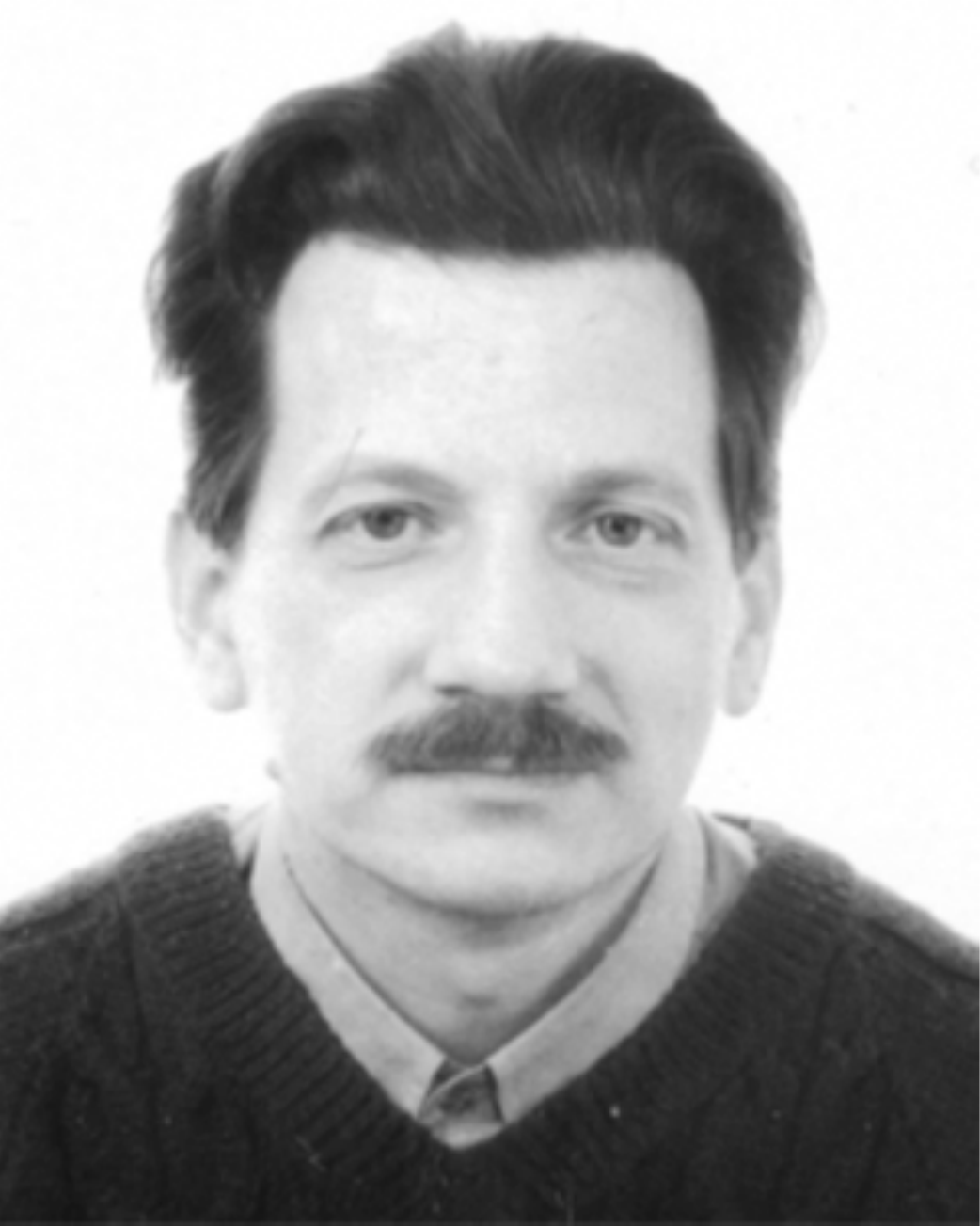}}]
{Alexey S. Matveev} was born in Leningrad, Russia, in 1954. He received the M.S. and Ph.D. degrees in applied mathematics and engineering cybernetics both from the Leningrad University, St. Petersburg, Russia, in 1976 and 1980, respectively. 

He is currently a Professor in the Department of Mathematics and Mechanics, Saint Petersburg University. His research interests include control over communication networks, hybrid dynamical systems, and navigation and control of mobile robots.
\end{IEEEbiography}

\begin{IEEEbiography}
[{\includegraphics[width=1in,height=1.25in,clip,keepaspectratio]{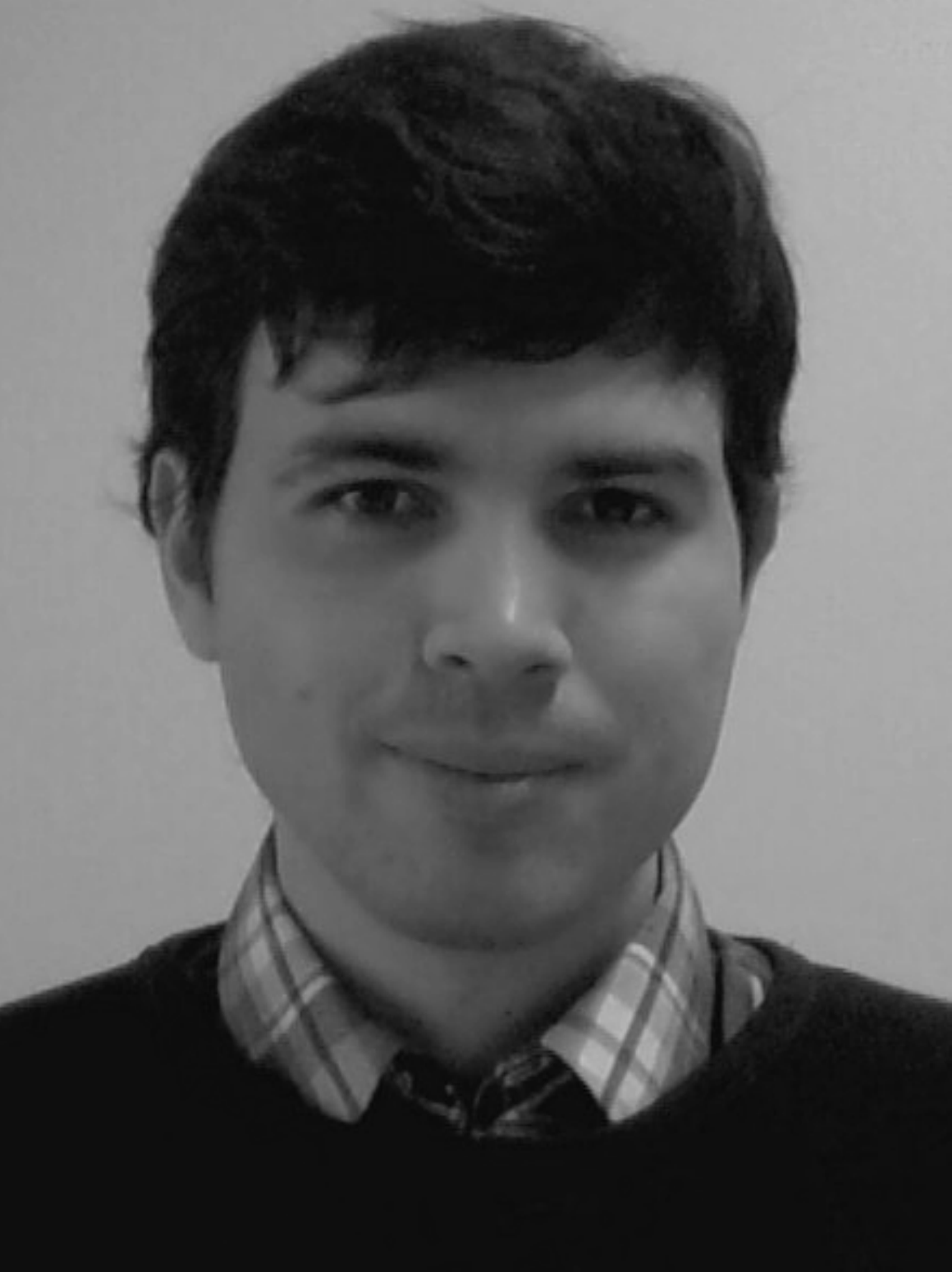}}]
{Juan E. Machado}
received the B.Sc. degree in electro\-mechanical engineering in 2012 from Instituto Tecnol\'{o}gico de La Paz, La Paz, M\'{e}xico and the M.Sc. degree in applied mathematics in 2015 from Centro de Investigaci\'{on} en Matem\'{a}ticas, Gua\-na\-jua\-to, M\'{e}xico. 

Currently, he is a Ph.D student at Universit\'{e} Paris Sud - Centrale Sup\'{e}lec, Gif-Sur-Yvette, France. His interests include modeling and control of electromechanical systems.
\end{IEEEbiography}

\begin{IEEEbiography}
[{\includegraphics[width=1in,height=1.25in,clip,keepaspectratio]{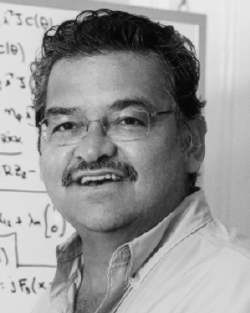}}]
{Romeo Ortega}
(S'81, M'85, SM'98, F'99) was born in Mexico. He obtained his BSc in
Electrical and Mechanical Engineering from the National University of
Mexico, Master of Engineering from Polytechnical Institute of
Leningrad, USSR, and the Docteur D`Etat from the Polytechnical
Institute of Grenoble, France in 1974, 1978 and 1984 respectively.

He then joined the National University of Mexico, where he worked until
1989. He was a Visiting Professor at the University of Illinois in
1987-88 and at the McGill University in 1991-1992, and a Fellow of the
Japan Society for Promotion of Science in 1990-1991.
He has been a member of the French National Researcher Council (CNRS) since
June 1992.  Currently he is in the Laboratoire de Signaux et Systemes (SUPELEC)
 in Gif--sur--Yvette.  His research interests are in the fields of
nonlinear and adaptive control, with special emphasis on applications. 

Dr Ortega has published  three books and more than 290 scientific papers in international journals, with an h-index of 79. He
has supervised more than 30 PhD thesis. He  has served as chairman in several IFAC and IEEE committees and participated in various editorial boards.
\end{IEEEbiography}

\begin{IEEEbiography}
[{\includegraphics[width=1in,height=1.25in,clip,keepaspectratio]{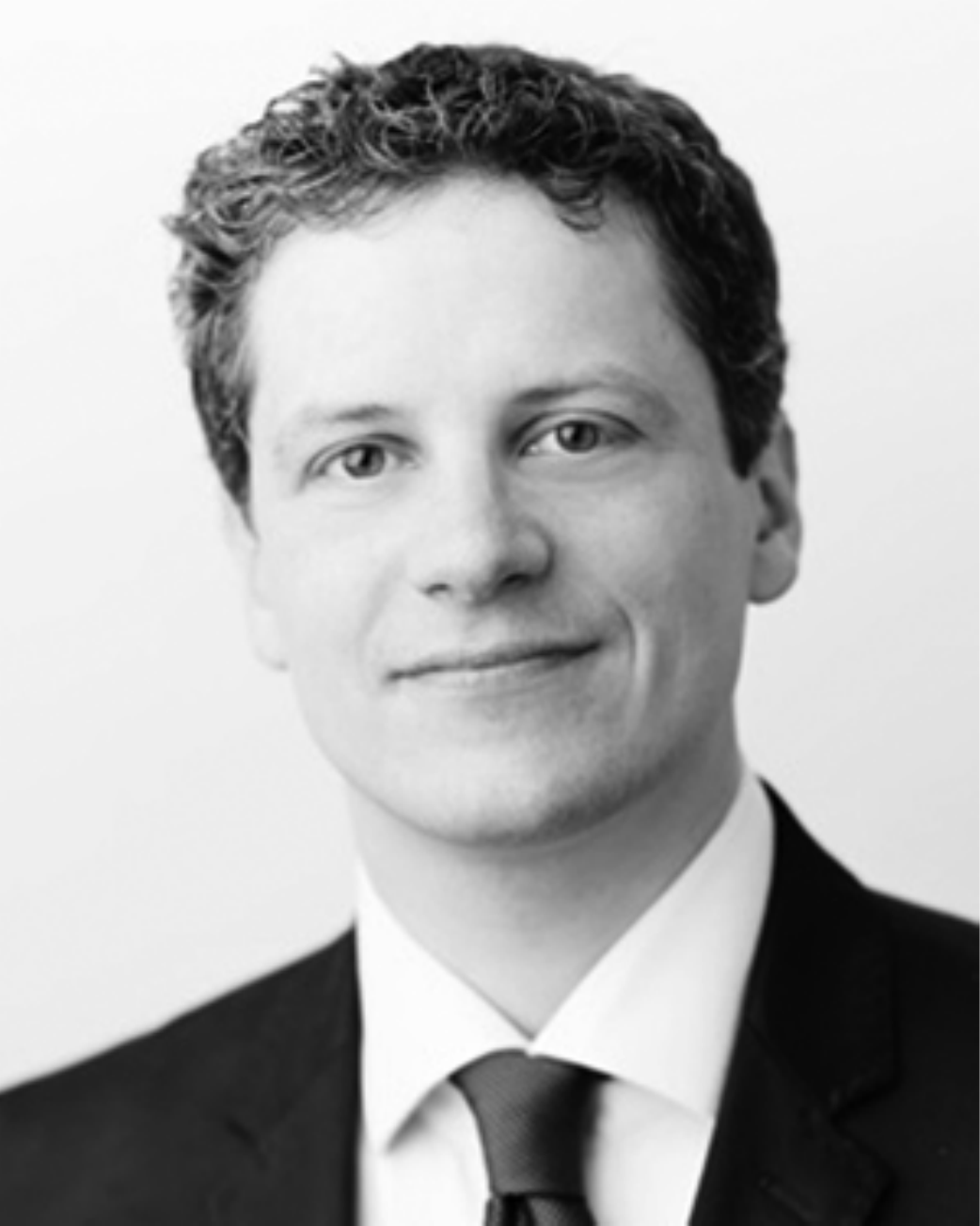}}]
{Johannes Schiffer} received the Diploma degree in engineering cybernetics from the University of Stuttgart, Stuttgart, Germany, in 2009 and the Ph.D. degree (Dr.-Ing.) in electrical engineering from Technische Universität (TU) Berlin, Berlin, Germany, in 2015.

He currently holds the chair of Control Systems and Network Control Technology at Brandenburgische Technische Universität Cottbus-Senftenberg, Cottbus, Germany. Prior to that, he has held appointments as Lecturer (Assistant Professor) at the School of Electronic and Electrical Engineering, University of Leeds, Leeds, U.K. and as Research Associate in the Control Systems Group and at the Chair of Sustainable Electric Networks and Sources of Energy both at TU Berlin. In 2017 he and his co-workers received the Automatica Paper Prize over the years 2014-2016. His current research interests include distributed control and analysis of complex networks with application to microgrids and power systems.
\end{IEEEbiography}

\begin{IEEEbiography}
[{\includegraphics[width=1in,height=1.25in,clip,keepaspectratio]{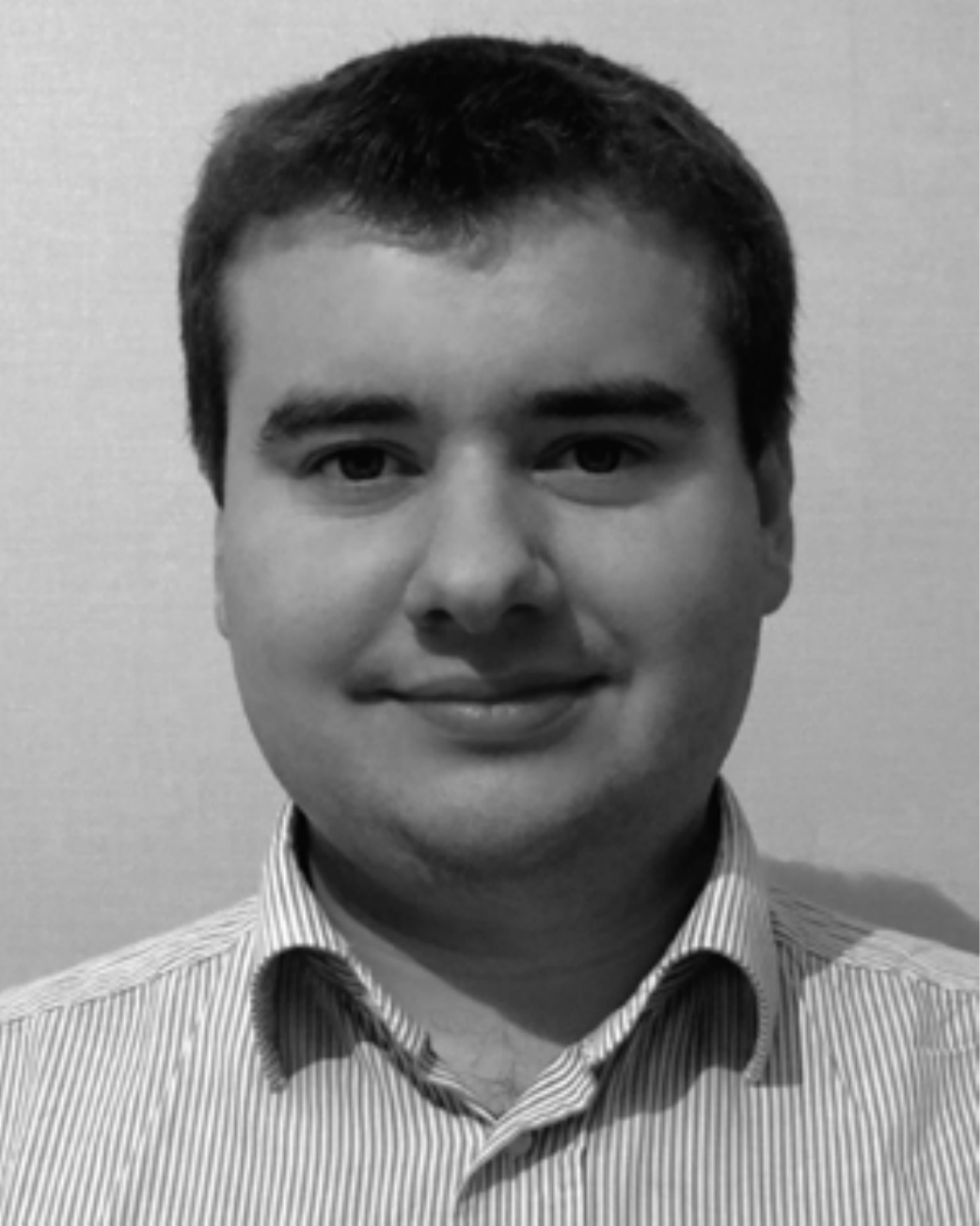}}]
{Anton Pyrkin} (M'11) was born in Zaozerniy, USSR, in 1985. He received the B.S. degree in 2006, the M.S. degree in 2008, the Ph.D. degree in 2010, and the Doctor of Science (habilitation thesis) degree in system analysis, data processing and control (in technical systems) in 2015, all from ITMO University, St. Petersburg, Russia.

He is a leading Scientist and Docent at the Department of Control Systems and Informatics, ITMO University. He has created the company “Robotronica Ltd.” main purpose of which is the research, developing, and assembling the mechatronic and robotic models of real technical plants for experimental approval of designed control systems. He is a coauthor of more than 100 publications in science journals and proceedings of conferences. His research interests include adaptive and robust control, frequency estimation, disturbance cancellation, time-delay systems, nonlinear systems, mechatronic and robotic systems, and autopilot and dynamic position systems for vessels.

He is a Member of the International Public Association Academy of Navigation and Motion Control.
\end{IEEEbiography}

\end{document}